\PassOptionsToPackage{pdfpagelabels=false}{hyperref}
\documentclass[useAMS,usedcolumn,usenatbib]{mnras}

\usepackage{mathptmx}
\usepackage{txfonts}

\usepackage[T1]{fontenc}
\usepackage{ae,aecompl}

\usepackage{graphicx}	% Including figure files
\usepackage{longtable} 
\usepackage{pdflscape} 
\usepackage{footnote}

\newcommand{\isotope}[2]{${}^{#1}$#2}
\newcommand{\msun}{\mbox{$\mathrm{M_{\odot}}$}}
\newcommand{\msunb}{\mbox{$\mathrm{M_{\odot}}$} }

\newcommand{\gcc}{\mbox {{\rm g~cm$^{-3}$}}}

\newcommand{\gccb}{\mbox {{\rm g~cm$^{-3}$}} }

\newcommand{\molg}{\mbox {{\rm mol~g$^{-1}$}}}

\newcommand{\ofusb}{$^{16}\mathrm{O}+^{16}\mathrm{O}$}
\newcommand{\cfusb}{$^{12}\mathrm{C}+^{12}\mathrm{C}$}
\newcommand{\cofusb}{$^{12}\mathrm{C}+^{16}\mathrm{O}$}

\newcommand{\nifsx}{$^{56}$Ni}
\newcommand{\nifsv}{$^{57}$Ni}
\newcommand{\feff}{$^{55}$Fe}
\newcommand{\nifsxb}{$^{56}$Ni }
\newcommand{\nifsvb}{$^{57}$Ni }
\newcommand{\feffb}{$^{55}$Fe }

\title[Post-processing accuracy]{The accuracy of post-processed nucleosynthesis}

\author[Bravo]{
Eduardo Bravo\thanks{E-mail: eduardo.bravo@upc.edu} 
\\
E.T.S. Arquitectura del Vall\`es, Universitat Polit\`ecnica de Catalunya, Carrer Pere Serra  
1-15, 08173 Sant Cugat del Vall\`es, Spain
}

\date{Accepted 2020 March 27. Received 2020 March 26; in original form 2020 January 14}

\pubyear{2020}

\begin{document}
\label{firstpage}
\pagerange{\pageref{firstpage}--\pageref{lastpage}}
\maketitle

\begin{abstract}
The computational requirements posed by multi-dimensional simulations of type Ia supernovae make it difficult to incorporate complex nuclear networks to follow the release of nuclear energy along with the propagation of the flame. Instead, these codes usually model the flame and use simplified nuclear kinetics, with the goal of determining a sufficiently accurate rate of nuclear energy generation and, afterwards, post-processing the thermodynamic trajectories with a large nuclear network to obtain more reliable nuclear yields. In this work, I study the performance of simplified nuclear networks with respect to reproduction of the 
nuclear yields obtained with a one-dimensional supernova code equipped with a large nuclear network. 
I start by defining a strategy to follow the properties of matter in nuclear statistical equilibrium (NSE). 
I propose to use published tables of NSE properties, together with a careful interpolation routine.
Short networks (iso7 and 13$\alpha$) are able to give an accurate yield of \nifsx, after post-processing, but can fail by order of magnitude in predicting the ejected mass of even mildly abundant species ($>10^{-3}$~\msun). A network of 21 species reproduces the nucleosynthesis of the Chandrasekhar and sub-Chandrasekhar explosions 
studied here with average errors better than 20\% for the whole set of stable elements and isotopes followed in the models.
\end{abstract}

\begin{keywords}
nuclear reactions, nucleosynthesis, abundances -- supernovae: general -- white dwarfs
\end{keywords}

\section{Introduction}\label{s:intro}

Type Ia supernovae (SNIa) are the result of the thermonuclear disruption of carbon-oxygen white dwarf (WD) stars, due to their destabilization by a companion star in a binary or a ternary system \citep{2000hil,2011how,2012kaz,2014mao}. The spectra and light curves of SNIa can be recorded starting shortly after the explosion \citep{2013zhe,2017hos,2018mie} and continuing until a few years later \citep{2018gra,2018jac,2018mag}, if the luminosity and the distance to the event allow it. Besides providing valuable insights into the systematics of 
SNIa \citep[][to cite only a few]{1993bran,1997fil,2002poz,2005how,2006jam,2008ars,2009bran},
these data allow researchers to constrain the properties of the explosion, from fundamental parameters, such as the ejected mass, explosion energy, and synthesized mass of \nifsxb \citep{2006stt,2014sca}, to second-order details, such as the ejected amounts of other radioactive isotopes, notably \nifsvb and \feffb \citep{2016gra,2017dim,2017sha,2018yan,2019li}. 
Whereas observational data are usually reported together with their error bars, there is scarce knowledge of the effects of the diverse sources of uncertainty related to supernova models. Indeed, \citet{2012bra,2013pkh,2019brab} proved that the nucleosynthesis of one-dimensional SNIa models is robust with respect to variations in individual reaction rates, at least during the explosion phase.

Currently, two competing models may account for the bulk of SNIa. In one model, the exploding WD is close to the Chandrasekhar-mass limit and the disruption is total; therefore, the mass of the ejecta is fixed a priori. The most accepted explosion mechanism of massive WDs is delayed detonation 
\citep[DDT,][]{1991kho,2019pol},
in which the thermonuclear burning wave propagates subsonically at first and turns into a detonation later, typically one or two seconds after thermal runaway. In the other model, the mass of the exploding WD is substantially less than the Chandrasekhar limit, the burning wave propagates as a detonation from the very beginning, and the instability leading to the detonation may be a consequence of the burning of a thin helium layer, accumulated on top of the WD after accretion from a secondary star, or it may be due to the merging or collision with another degenerate star 
\citep[e.g. a second WD;][]{1994woob,2010woo,2010sim,2013kus,2018she}. 

Although there is not consensus about the level of asymmetry involved in SNIa explosions, most polarization measurements are indicative of small deviations from spherical symmetry \citep{1996wan,2001how,2010mau,2010maeb,2013mau}. One-dimensional models are able to account for the properties of SNIa in the visible \citep{1996hoe,1997nug,2011tan,2013blo,2017hoe} and gamma bands \citep[][see \citealp{2014die,2016ise} for a different view]{2014chu,2015chu}, and those of their remnants in the X-ray and radio bands \citep[e.g.][]{2006bad,2011lop,2018mar}. However, to understand SNIa it is necessary to simulate the explosion in three dimensions, either to account for hydrodynamical instabilities and turbulence \citep[e.g.][]{2004ple,2006rop,2006bra,2007kasb}, or to address the inherent asymmetrical configuration of colliding or merging WDs. 

The computational requirements posed by three-dimensional hydrodynamics make it difficult to incorporate complex nuclear networks to follow the release of nuclear energy along with the propagation of the flame. Usually, multi-dimensional supernova codes need to model the flame by making use of simplified nuclear kinetics, with the goals of giving an accurate rate of nuclear energy generation and computing the explosion in a reasonable time. Afterwards, the thermodynamic trajectories of the integration nodes can be post-processed with the aid of a large nuclear network to obtain more reliable values of the supernova yields \citep[e.g.][]{1986thi,2010bra,2016tow,2017leu}.  
Although, recently, several groups have designed algorithms to incorporate large nuclear networks into multi-dimensional models of SNIa \citep{2015pap,2019kusb}, most research still follows the post-processing approach.

In this work, I address the question of the performance of simplified nuclear networks with respect to the reproduction of the correct nuclear yields\footnote{In the present work, by correct nuclear yields I mean those obtained with the same supernova code without the simplifications described in the text.}. After a brief explanation of the methodology used (Sect.~\ref{s:2}), the first point I treat is the definition of a strategy to follow the properties of matter in a state of nuclear statistical equilibrium (NSE, Sect.~\ref{s:3}). In the following section, I test several simplified nuclear networks for the accuracy of the nucleosynthesis obtained through post-processing (Sect.~\ref{s:4}). 
Two more sections address the performance of simplified nuclear networks for high-metallicity WD progenitors and the effect of different criteria for switching on and off NSE routines. A final section is dedicated to the conclusions of the present work.

\section{Methodology}\label{s:2}

The simulations presented in this work are performed with the same supernova code described in \citet{2019bra}, where extensive details of the method of computation can be found. The code integrates the hydrodynamic evolution using a large nuclear network, solves the Saha equations for NSE, when applicable, and calculates the neutronization rate and associated neutrino energy losses at each time step, computing the weak interaction rates using the NSE composition. The hydrodynamics is followed in one dimension, assuming spherical symmetry, but the nucleosynthetic processes through which matter passes are 
qualitatively\footnote{The precise thermodynamic histories of mass shells may be a function of the dimensionality of the model.}
the same as in any three-dimensional SNIa simulation, hence it is appropriate to assess the accuracy of the computed nucleosynthesis. Furthermore, the supernova models generated by this code have been compared successfully with observed optical spectra (S. Blondin, private communication), gamma-ray emission from SN2014J \citep{2014chu,2015chu,2016ise} and the X-ray spectra of supernova remnants \citep{2006bad}.
 
The default nuclear network used here is the same as in \citet{2012bra} and can be found in column BM-P in Table~\ref{t:1}. This network includes all stable isotopes up to molybdenum, but the code allows to define different nuclear networks by simply listing the species to be followed. The nuclear reactions linking these species are included automatically in the network, together with a basic set formed by the fusion reactions triple-$\alpha$, \cfusb, \ofusb, and \cofusb. All reaction rates are taken from the JINA REACLIB\footnote{http://groups.nscl.msu.edu/jina/reaclib/db/.} compilation \citep{2010cyb},
in the version of November 6, 2008. 

As references, I have selected two explosion models suitable for normal-luminosity SNIa, characterized by
an ejected mass of \nifsxb about $M(^{56}\mathrm{Ni})\sim0.5 - 0.7~\msunb$. The first model, sub-$M_\mathrm{Ch}$, is a central detonation of a sub-Chandrasekhar WD of mass $M_\mathrm{WD} = 1.06~\msun$ \citep[model 1p06\_Z9e-3\_std in][]{2019bra}. The second one, $M_\mathrm{Ch}$, is a delayed detonation of a massive WD with central density $\rho_\mathrm{c} = 3\times10^9$~\gccb and a deflagration-to-detonation transition density $\rho_\mathrm{DDT}=2.4\times10^7$~\gccb \citep[model ddt2p4\_Z9e-3\_std in][]{2019bra}. Both models assume an initial composition made of equal masses of \isotope{12}{C} and \isotope{16}{O}, contaminated with \isotope{22}{Ne} as appropriate for progenitor metallicity $0.009$ and other metals from sodium to indium in solar proportions with respect to \isotope{22}{Ne}.

\begin{table}
%\begin{minipage}{126mm}
 \caption{Nuclear networks for the convergence study.}
 \label{t:1}
 \begin{tabular}{lccc}
 \hline\hline
\noalign{\smallskip}
  & BM-P & netAKh & nse7 \\
$Z$ & $A_\mathrm{min}-A_\mathrm{max}$ & $A_\mathrm{min}-A_\mathrm{max}$ & $A_\mathrm{min}-A_\mathrm{max}$ \\
 \hline\hline
\noalign{\smallskip}
 n & 1 - 1 & 1 - 1 & 1 - 1 \\
 H & 1 - 4 & 1 - 1 & 1 - 3 \\
He & 3 - 9 & 4 - 4 & 3 - 6 \\
Li & 4 - 11 & - & 6 - 7 \\
Be & 6 - 14 & - & 7 - 10 \\
 B & 7 - 17 & - & 10 - 11 \\
 C & 8 - 20 & 12 - 13 & 11 - 14 \\
 N & 10 - 21 & 13 - 13 & 13 - 15 \\
 O & 12 - 23 & 16 - 16 & 15 - 18 \\
 F & 14 - 25 & - & 17 - 19 \\
Ne & 16 - 27 & 20 - 22 & 19 - 23 \\
Na & 18 - 34 & 23 - 23 & 21 - 25 \\
Mg & 20 - 35 & 23 - 26 & 23 - 28 \\
Al & 22 - 36 & 27 - 27 & 25 - 30 \\
Si & 24 - 38 & 27 - 32 & 27 - 33 \\
 P & 26 - 40 & 30 - 33 & 29 - 35 \\
 S & 28 - 42 & 31 - 36 & 30 - 37 \\
Cl & 30 - 44 & 35 - 37 & 32 - 39 \\
Ar & 32 - 46 & 36 - 41 & 34 - 42 \\
 K & 34 - 49 & 39 - 43 & 37 - 45 \\
Ca & 36 - 51 & 40 - 46 & 38 - 48 \\
Sc & 38 - 52 & 41 - 47 & 41 - 51 \\
Ti & 40 - 54 & 43 - 50 & 43 - 53 \\
 V & 42 - 56 & 45 - 52 & 45 - 55 \\
Cr & 44 - 58 & 47 - 56 & 47 - 57 \\
Mn & 46 - 60 & 49 - 60 & 49 - 59 \\
Fe & 49 - 63 & 51 - 62 & 50 - 62 \\
Co & 51 - 65 & 53 - 61 & 52 - 64 \\
Ni & 53 - 69 & 56 - 64 & 54 - 66 \\
Cu & 55 - 71 & 57 - 65 & 56 - 68 \\
Zn & 57 - 78 & 59 - 66 & 58 - 70 \\
Ga & 61 - 81 & - & 61 - 72 \\
Ge & 63 - 83 & - & 64 - 74 \\
As & 65 - 85 & - & 69 - 75 \\
Se & 67 - 87 & - & 75 - 75 \\
Br & 69 - 90 & - & - \\
Kr & 71 - 93 & - & - \\
Rb & 73 - 99 & - & - \\
Sr & 77 - 100 & - & - \\
 Y & 79 - 101 & - & - \\
Zr & 81 - 101 & - & - \\
Nb & 85 - 101 & - & - \\
Mo & 87 - 101 & - & - \\
Tc & 89 - 101 & - & - \\
Ru & 91 - 101 & - & - \\
Rh & 93 - 101 & - & - \\
Pd & 95 - 101 & - & - \\
Ag & 97 - 101 & - & - \\
Cd & 99 - 101 & - & - \\
In & 101 - 101 & - & - \\
\hline\hline
 \end{tabular}
\end{table}

\begin{table*}
\caption{Nucleosynthetic indicators.} 
\label{t:2} 
\centering 
\begin{tabular}{llcccccccccccccc}
\hline\hline
\noalign{\smallskip}
 & & $a_1$ & $a_2$ & $b_1$ & $b_2$ & $c_1$ & $c_2$ & $c_3$ & $c_4$ & $d_1$ & $d_2$ & $d_3$ & $e_1$ & $e_2$ & $e_3$ \\
 &  & (\%) & (\%) & (\%) & (\%) & (\%) &(\%)  & (\%) & (\%) & (\%) &(\%)  & (\%) & (\%) & (\%) & (\%) \\
\hline
\noalign{\smallskip}
\multicolumn{2}{l}{Convergence study} & \multicolumn{14}{c}{} \\
Network & Model & \multicolumn{14}{c}{} \\
netAKh & $M_\mathrm{Ch}$ & 0.3 & 0.2 & 0.4 & 0.0 & 1.5 & 1.5 & 0.3 & 0.4 & 1.5 & 6.1 & 1.3 & 1.5 & 3.9 & 6.1 \\
nse7   & $M_\mathrm{Ch}$ & 0.0 & 0.0 & 0.0 & 0.0 & 0.3 & 0.3 & 0.0 & 0.0 & 0.2 & 0.7 & 1.4 & 0.2 & 0.5 & 1.4 \\
netAKh & sub-$M_\mathrm{Ch}$   & 0.3 & 0.0 & 0.3 & 0.3 & 0.7 & 0.8 & 0.2 & 0.2 & 1.2 & 1.8 & 0.8 & 1.2 & 1.8 & 1.6 \\
nse7   & sub-$M_\mathrm{Ch}$   & 0.0 & 0.0 & 0.0 & 0.0 & 0.3 & 0.3 & 0.1 & 0.1 & 0.0 & 1.4 & 0.3 & 0.0 & 1.3 & 1.4 \\
\hline
\noalign{\smallskip}
\multicolumn{2}{l}{NSE table interpolation} & \multicolumn{14}{c}{} \\
Interpolator & Model & \multicolumn{14}{c}{} \\
linlog & $M_\mathrm{Ch}$ & 0.2 & 0.6 & 11 & 7.6 & 4.3 & 23 & 0.9 & 4.4 & 7.4 & 12 & 0.8 & 22 & 100 & 360 \\
linlin  & $M_\mathrm{Ch}$ & 0.0 & 0.0 & 0.1 & 0.0 & 1.0 & 8.3 & 0.2 & 1.5 & 0.9 & 4.7 & 0.3 & 8.2 & 32 & 82 \\
loglog & $M_\mathrm{Ch}$ & 0.2 & 0.6 & 13 & 7.1 & 4.3 & 23 & 0.9 & 4.4 & 7.4 & 12 & 0.8 & 22 & 101 & 355 \\
poly & $M_\mathrm{Ch}$ & 0.1 & 0.5 & 13 & 7.4 & 4.8 & 28 & 0.9 & 5.0 & 7.9 & 13 & 1.9 & 24 & 120 & 490 \\
spline & $M_\mathrm{Ch}$ & 0.0 & 0.0 & 0.1 & 0.0 & 0.1 & 0.2 & 0.0 & 0.1 & 0.1 & 0.2 & 0.4 & 0.6 & 0.7 & 0.8 \\
pe$\nu$n & $M_\mathrm{Ch}$ & 0.0 & 0.1 & 0.3 & 0.1 & 0.2 & 0.6 & 0.1 & 0.1 & 0.3 & 0.3 & 1.4 & 0.8 & 1.1 & 4.2 \\
den$\times10$ & $M_\mathrm{Ch}$ & 0.0 & 0.1 & 0.4 & 0.0 & 0.2 & 0.5 & 0.1 & 0.1 & 0.3 & 0.3 & 0.6 & 0.8 & 1.0 & 3.1 \\
\hline
\noalign{\smallskip}
\multicolumn{2}{l}{Simplified networks} & \multicolumn{14}{c}{} \\
Network & Model & \multicolumn{14}{c}{} \\
iso7 & $M_\mathrm{Ch}$ & 0.5 & 9.4 & 86 & 29 & 58 & 120 & 12 & 23 & 64 & 620 & 14 & 310 & 1500 & 5900 \\
$13\alpha$ & $M_\mathrm{Ch}$ & 3.7 & 4.3 & 70 & 27 & 43 & 93 & 7.4 & 17 & 49 & 380 & 8.8 & 220 & 910 & 3100 \\
net21 & $M_\mathrm{Ch}$ & 4.3 & 3.2 & 6.4 & 6.5 & 8.9 & 8.4 & 3.5 & 4.1 & 16 & 17 & 12 & 17 & 18 & 14 \\
iso7 & sub-$M_\mathrm{Ch}$ & 0.8 & 6.1 & 26 & 31 & 33 & 43 & 11 & 15 & 52 & 140 & 9.8 & 92 & 170 & 320 \\
$13\alpha$ & sub-$M_\mathrm{Ch}$ & 1.2 & 3.9 & 21 & 27 & 23 & 30 & 7.5 & 11 & 40 & 93 & 7.4 & 66 & 110 & 180 \\
net21 & sub-$M_\mathrm{Ch}$ & 1.4 & 1.1 & 0.0 & 2.4 & 3.6 & 3.7 & 1.4 & 1.6 & 5.0 & 9.4 & 4.0 & 5.0 & 6.1 & 9.5 \\
\hline
\noalign{\smallskip}
\multicolumn{2}{l}{High metallicity progenitor} & \multicolumn{14}{c}{} \\
Network & Model & \multicolumn{14}{c}{} \\
net21 & $M_\mathrm{Ch}$ & 3.8 & 4.5 & 12 & 4.4 & 11 & 12 & 3.7 & 7.4 & 19 & 19 & 17 & 21 & 20 & 31 \\
net23 & $M_\mathrm{Ch}$ & 6.0 & 2.5 & 2.8 & 9.0 & 15 & 18 & 4.2 & 7.6 & 28 & 29 & 22 & 28 & 69 & 26 \\
net21 & sub-$M_\mathrm{Ch}$ & 2.4 & 5.9 & 9.4 & 6.2 & 16 & 18 & 6.8 & 9.8 & 20 & 29 & 22 & 25 & 30 & 32 \\
net23 & sub-$M_\mathrm{Ch}$ & 2.9 & 0.1 & 1.0 & 4.5 & 5.5 & 6.5 & 1.7 & 3.1 & 7.6 & 11 & 7.0 & 10 & 13 & 10 \\
\hline
\noalign{\smallskip}
\multicolumn{2}{l}{Transition NSE $\leftrightarrows$ net21} & \multicolumn{14}{c}{} \\
Condition & Model & \multicolumn{14}{c}{} \\
$T_\mathrm{NSE}=5.5\times10^9$~K & $M_\mathrm{Ch}$ & 4.3 & 3.2 & 6.4 & 6.5 & 8.9 & 8.4 & 3.5 & 4.1 & 16 & 17 & 12 & 17 & 18 & 14 \\
$T_\mathrm{NSE}=5.0\times10^9$~K & $M_\mathrm{Ch}$ & 3.5 & 1.6 & 10 & 7.0 & 5.2 & 5.3 & 1.9 & 2.4 & 8.4 & 7.9 & 8.3 & 9.6 & 14 & 8.3 \\
$T_\mathrm{out}=4\times10^9$~K  & $M_\mathrm{Ch}$ & 3.3 & 0.5 & 14 & 8.0 & 4.3 & 5.1 & 1.2 & 1.9 & 9.7 & 9.7 & 4.8 & 12 & 20 & 16 \\
$T_\mathrm{out}=3\times10^9$~K  & $M_\mathrm{Ch}$ & 3.3 & 1.1 & 1.4 & 1.6 & 19 & 40 & 2.9 & 8.9 & 21 & 100 & 5.1 & 70 & 260 & 530 \\
$\rho_\mathrm{NSE0}=10^8$~\gcc  & $M_\mathrm{Ch}$ & 4.4 & 3.3 & 4.1 & 4.6 & 8.8 & 8.3 & 3.6 & 4.2 & 17 & 17 & 12 & 17 & 18 & 15 \\
$\rho_\mathrm{NSE0}=4\times10^7$~\gcc  & $M_\mathrm{Ch}$ & 4.3 & 3.2 & 6.7 & 6.3 & 8.9 & 8.4 & 3.5 & 4.1 & 17 & 17 & 12 & 17 & 19 & 14 \\
\hline\hline
\end{tabular}
\end{table*}

As a first test of convergence of the supernova code with respect to the size of the nuclear network, I have re-computed models $M_\mathrm{Ch}$ and sub-$M_\mathrm{Ch}$ with two different nuclear networks, both of them sufficiently large to give accurate nuclear energy generation rates. The first network (netAKh) is that employed in the SNIa models computed by Alexei Khokhlov and reported in \citet{2013blo,2017blo}. It uses 144 isotopes, including all stable isotopes between neon and copper, with the exception of \isotope{48}{Ca}. The second network (nse7) includes 260 isotopes and was introduced by \citet{2019kus} and used in their study of the structure of detonation waves in SNIa. It includes all stable isotopes up to arsenic, with the exception of \isotope{76}{Ge}. Table~\ref{t:1} shows both networks. 

To assess the accuracy of the nucleosynthetic yields I use a set of fourteen indicators, all of them expressed as per cent relative differences between the quantities obtained in a test model with respect to the results in the reference hydrodynamic model that uses the default reaction network. Therefore, I compare sub-Chandrasekhar models with model 1p06\_Z9e-3\_std, and Chandrasekhar-mass models with ddt2p4\_Z9e-3\_std. The indicators are as follows:
\begin{itemize}
 \item the discrepancy of the final kinetic energy, $a_1=\Delta K/K$, 
 \item the discrepancy of the ejected mass of \nifsx, \mbox{$a_2=\Delta M(^{56}\mathrm{Ni})/M(^{56}\mathrm{Ni})$}, 
 \item the discrepancy of the ejected mass of \nifsvb, \mbox{$b_1=\Delta M(^{57}\mathrm{Ni})/M(^{57}\mathrm{Ni})$}, 
 \item the discrepancy of the ejected mass of \feff, \mbox{$b_2=\Delta M(^{55}\mathrm{Fe})/M(^{55}\mathrm{Fe})$},
 \item a measure of the discrepancy, $c_1$, based on the average of the squared deviations of the logarithm of the ejected mass of the elements, 
\begin{equation}
 \sigma_\mathrm{log,ele}=\sqrt{\frac{1}{N_\mathrm{ele}}\sum\log^2\left[\frac{M'(Z)}{M(Z)}\right]}\,,
\end{equation}
\begin{equation}
 c_1=10^{\sigma_\mathrm{log,ele}} - 1\,,
\end{equation}
 \item a measure of the discrepancy, $c_2$, based on the average of the squared deviations of the logarithm of the ejected mass of the isotopes, 
\begin{equation}
 \sigma_\mathrm{log,iso}=\sqrt{\frac{1}{N_\mathrm{iso}}\sum\log^2\left[\frac{M'(^{A}Z)}{M(^{A}Z)}\right]}\,,
\end{equation}
\begin{equation}
 c_2=10^{\sigma_\mathrm{log,iso}} - 1\,,
\end{equation}
 \item a measure of the discrepancy, $c_3$, based on a weighted average of the squared deviations of the logarithm of the ejected mass of the elements, 
\begin{equation}
 \sigma_\mathrm{wm,ele}=\sqrt{\sum\omega(Z)\log^2\left[\frac{M'(Z)}{M(Z)}\right]}\,,
\end{equation}
\begin{equation}
 c_3=10^{\sigma_\mathrm{wm,ele}} - 1\,,
\end{equation}
 \item a measure of the discrepancy, $c_4$, based on a weighted average of the squared deviations of the logarithm of the ejected mass of the isotopes, 
\begin{equation}
 \sigma_\mathrm{wm,iso}=\sqrt{\sum\omega(^{A}Z)\log^2\left[\frac{M'(^{A}Z)}{M(^{A}Z)}\right]}\,,
\end{equation}
\begin{equation}
 c_4 = 10^{\sigma_\mathrm{wm,iso}} - 1\,,
\end{equation}
 \item the maximum relative discrepancy, $d_1$, in the mass of the elements with final yield $M(Z)\ge10^{-3}$~\msun,
 \item the maximum relative discrepancy, $d_2$, in the mass of the elements with final yield in the range $10^{-3}>M(Z)\ge10^{-6}$~\msun,
 \item the maximum relative discrepancy, $d_3$, in the mass of the elements with final yield in the range $10^{-6}>M(Z)\ge10^{-12}$~\msun,
 \item the maximum relative discrepancy, $e_1$, in the mass of the isotopes with final yield $M(^{A}Z)\ge10^{-3}$~\msun,
 \item the maximum relative discrepancy, $e_2$, in the mass of the isotopes with final yield in the range $10^{-3}>M(^{A}Z)\ge10^{-6}$~\msun, and
 \item the maximum relative discrepancy, $e_3$, in the mass of the isotopes with final yield in the range $10^{-6}>M(^{A}Z)\ge10^{-12}$~\msun.
\end{itemize}
The quantities $M(Z)$ and $M(^{A}Z)$ are the masses, in \msun, of element $Z$ and isotope $^{A}Z$ in the reference model, $M'$ stands for the same quantities for the test model, $N_\mathrm{ele}$ and $N_\mathrm{iso}$ are, respectively, the number of different elements and isotopes ejected and the weighting functions are defined as:
\begin{equation}
 \omega(Z)=\frac{1}{\left[\log^2M(Z)\right]\times\sum\left[1/\log^2M(Z)\right]}\,,
\end{equation}
\begin{equation}
 \omega(^{A}Z)=\frac{1}{\left[\log^2M(^{A}Z)\right]\times\sum\left[1/\log^2M(^{A}Z)\right]}\,.
\end{equation}
With these definitions, indicators $c_3$ and $c_4$ provide a measure of the mean deviation of the most abundant species, while indicators $c_1$ and $c_2$ give a measure of the deviation of the yields of all species. Indicators $a_2$, $b_1$, and $b_2$ are evaluated from the ejected masses of the isotopes 100~s after thermal runaway. All the indicators from $c_1$ to $e_3$ refer to the elemental or isotopic yields of isotopes between carbon and krypton after radioactive decays. 

For reference, the kinetic energy and masses of radioactive isotopes ejected in the two reference models are the following: in model ddt2p4\_Z9e-3\_std, $K=1.42\times10^{51}$~erg, $M(^{56}\mathrm{Ni})=0.685$~\msun, $M(^{57}\mathrm{Ni})=7.15\times10^{-3}$~\msunb and $M(^{55}\mathrm{Fe})=9.65\times10^{-3}$~\msun; in model 1p06\_Z9e-3\_std, $K=1.32\times10^{51}$~erg, $M(^{56}\mathrm{Ni})=0.664$~\msun, $M(^{57}\mathrm{Ni})=1.01\times10^{-2}$~\msunb and $M(^{55}\mathrm{Fe})=3.21\times10^{-3}$~\msun.

The thermodynamic trajectories obtained with the netAKh and nse7 networks have been fed to a post-processing nuclear code that uses the same network and reaction rates as in the BM-P network\footnote{The same strategy has been applied to all the calculations presented in the following sections.}. 
Table~\ref{t:2} show the results in the rows under the header ``Convergence study``. The agreement with the results of the hydrodynamic calculation using the default network is very satisfactory. The direct measure of the nuclear energy released, that is the final kinetic energy, is reproduced in all four calculations to better than 0.3\%. The yields of the radioactive isotopes are reproduced to within 0.4\% with both networks and, in particular, the yield of \nifsxb to better than 0.2\%. 

The nucleosynthesis of stable isotopes and elements also converges, where both netAKh and nse7 obtain similar ratings. The mean deviation of the most abundant elements and isotopes is $\le0.4$\%, while that representative of all ejected species lies in the range 0.3-1.5\%. Finally, the maximum deviation of elements and isotopes whose yield is larger than $10^{-3}$~\msunb is $\le1.5$\%, while that of the remaining elements and isotopes with yields $\ge10^{-12}$~\msunb is less than $\sim6$\%.

\section{Nuclear statistical equilibrium}\label{s:3}

One of the key ingredients of simulations of SNIa explosions is the treatment of NSE in matter burnt at high density. The composition of matter in NSE can be calculated by solving a set of Saha equilibrium equations linking the abundances of all isotopes to two arbitrarily chosen abundances (or combinations thereof), which play the role of independent variables, plus two closure relationships that account for the conservation of baryon number and the electrical neutrality of matter. The procedure is usually iterative, which makes it inefficient for a multi-dimensional hydrodynamic computation of a supernova explosion. Therefore, in this sort of simulation, it is usual to rely on interpolation of a table of NSE states, pre-computed on a net of density, $\rho$, temperature, $T$, and electron mole number, $Y_\mathrm{e}$, nodes, the denser the better. Usually, the table gives the main properties of matter in NSE, including nuclear binding energy, mean molar number, neutronization rate and neutrino energy loss rate.

One example of this kind of table of NSE properties is given in \citet{2009se2}. Recently, \citet{2019bra} reported that the final yields computed using their NSE table might disagree by order of magnitude for some isotopes with respect to those obtained computing the NSE state properties on the fly in the hydrodynamical calculation. The discrepancy did not affect significantly either the total energy release, i.e. the final kinetic energy, or the ejected mass of \nifsx, and was attributed to the interpolation procedure applied to obtain the NSE properties out of the $\rho$, $T$, and $Y_\mathrm{e}$ table nodes.

\begin{figure}
   \includegraphics[width=\columnwidth]{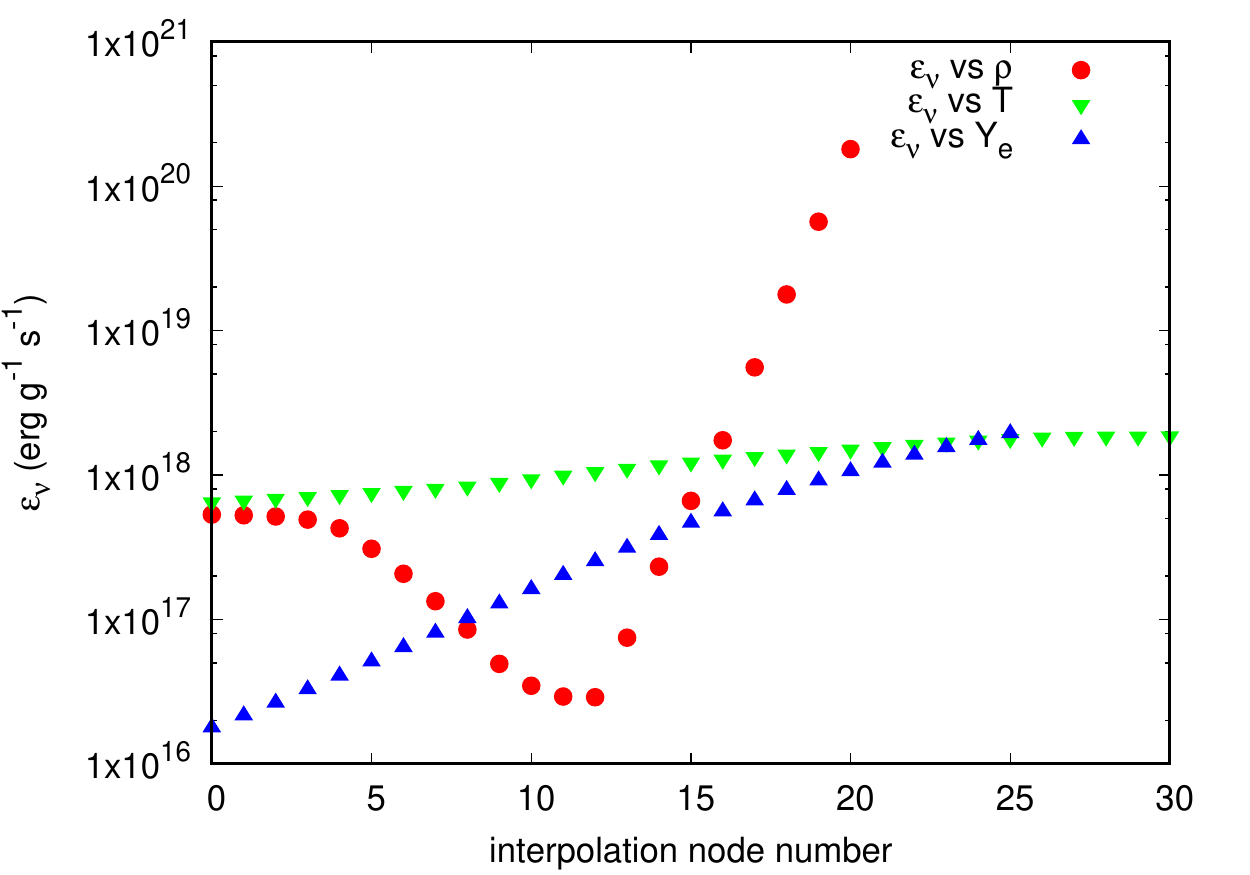}
\caption{Sample of variation of the neutrino energy loss-rate as function of density in the range $2\times10^5-2\times10^{10}$~\gccb (at fixed $T=9\times10^9$~K and $Y_\mathrm{e}=0.5$~mol~g$^{-1}$, red dots), temperature in the range $3\times10^9-1.05\times10^{10}$~K (at fixed $\rho=2\times10^9$~\gccb and $Y_\mathrm{e}=0.5$~mol~g$^{-1}$, green downward triangles) and electron mole number in the range $0.4400-0.5025$~mol~g$^{-1}$ (at fixed $\rho=2\times10^9$~\gccb and $T=9\times10^9$~K, blue upward triangles) in the interpolation nodes of the NSE table. 
}
\label{f:1}
\end{figure}

Here, I argue that the culprit for the discrepancy just mentioned is relying on the interpolation at the density nodes. To illustrate the situation, Fig.~\ref{f:1} shows the neutrino energy loss-rate, $\varepsilon_\nu$ in a sample of the \citet{2009se2} table nodes of $\rho$, $T$, and $Y_\mathrm{e}$ for typical values during a supernova explosion. While the energy-loss rate changes smoothly between consecutive temperature and electron mole number nodes, the dependence on density is more complex and the values of $\varepsilon_\nu$ change by four orders of magnitude between $\rho$ node numbers 12 and 20 ($\rho=2\times10^8$~\gccb and $\rho=2\times10^{10}$~\gcc). This huge change makes the results of the interpolation in density sensitive to the interpolation procedure. Indeed, the plot of $\varepsilon_\nu$ versus density in between nodes 12 and 20 in Fig.~\ref{f:1} is suggestive of a linear dependence between $\log\varepsilon_\nu$ and $\log\rho$ (the table nodes are equispaced in $\log\rho$).

To test the impact of the NSE table interpolation scheme I have chosen different interpolants and computed the difference in NSE properties with respect to those obtained by solving the NSE Saha equilibrium equations. The interpolants used and the designations given are the following:
\begin{itemize}
 \item linear interpolation of NSE properties, for example $\varepsilon_\nu$ or the neutronization rate $\dot{Y}_\mathrm{e}$, with respect to $\log\rho$ (linlog);
 \item linear interpolation with respect to $\rho$ (linlin);
 \item linear interpolation of $\log\dot{Y}_\mathrm{e}$ and $\log\varepsilon_\nu$ with respect to $\log\rho$ (loglog);
 \item third order polynomial of the NSE properties, for example $\dot{Y}_\mathrm{e}$, with respect to $\log\rho$ (poly);
 \item cubic spline fitting the NSE properties with respect to $\rho$ (spline);
 \item a physically motivated interpolation function (pe$\nu$n).
\end{itemize}
The last interpolator is motivated by the dominant role of protons in the neutronization rate of NSE matter in SNIa models \citep{1985ful,2000brc,2019brab}. Hence it seems natural to interpolate using the same function that describes the dependence of the p(e$^{-}$,$\nu$)n rate or the associated neutrino energy emission rate on density \citep{1985ful}. The effective log($ft$)-values characterizing electron captures by protons are almost constant, whereas the rate dependence on $\rho$ is given by the so-called modified phase space factor \citep[Eqs.~3 and 6 in][]{1985ful}, $I_\mathrm{e}$. The neutrino energy emission rate depends on $\rho$ through the appropriate phase-space factor \citep[Eq.~7 in][]{1985ful}, $J_\mathrm{e}^\nu$. Instead of computing the relativistic Fermi integrals that appear in the definition of these space factors, I calculate approximate values taking advantage of Eqs.~15 in \citet{1985ful}. 

\begin{figure*}
   \includegraphics[width=\textwidth]{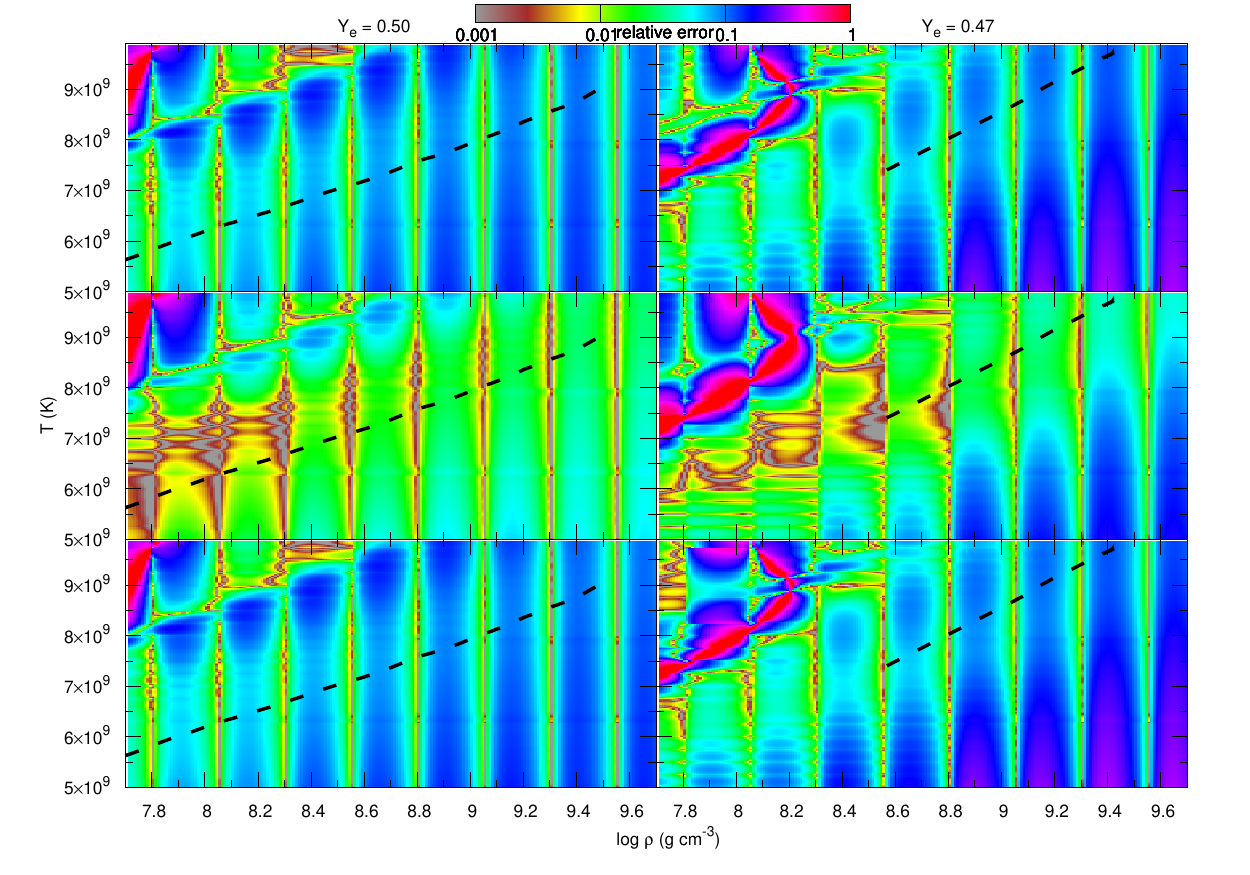}
\caption{Relative error between the exact neutronization rate, $\dot{Y}_\mathrm{e}$, and that computed by interpolation on a NSE table, for two values of the electron mole number, $Y_\mathrm{e}=0.50$~mol~g$^{-1}$ (left column) and $Y_\mathrm{e}=0.47$~mol~g$^{-1}$ (right column). The results are shown for different interpolation schemes, from top to bottom: linear interpolation of $\dot{Y}_\mathrm{e}$ versus $\log\rho$, linear interpolation of $\dot{Y}_\mathrm{e}$ versus $\rho$ and linear interpolation of $\log\dot{Y}_\mathrm{e}$ versus $\log\rho$. The relative error is colour coded according to the colour bar at the top of the plot. The dashed lines show the density and temperature at which the mass shells of model $M_\mathrm{Ch}$ start experiencing electron captures in NSE (left column) and at the time they reach an electron mole number $Y_\mathrm{e}=0.47$~mol~g$^{-1}$ (right column).
}
\label{f:2}
\end{figure*}

\begin{figure*}
   \includegraphics[width=\textwidth]{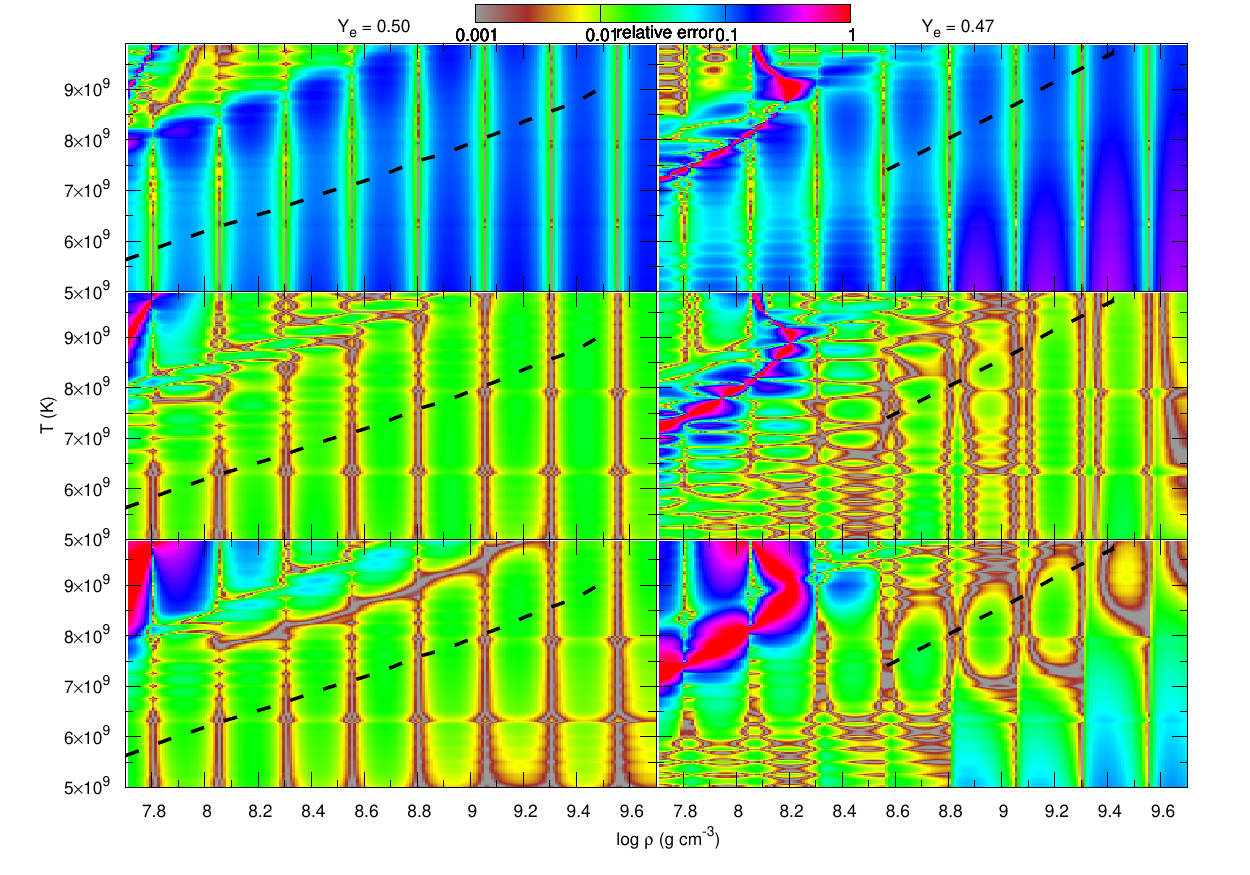}
\caption{Same as Fig.~\ref{f:2} but for different interpolation schemes, from top to bottom: third order polynomial of $\dot{Y}_\mathrm{e}$ versus $\log\rho$, cubic spline of $\dot{Y}_\mathrm{e}$ versus $\rho$ and pe$\nu$n.
}
\label{f:3}
\end{figure*}

Figures \ref{f:2} and \ref{f:3} show the relative error between $\dot{Y}_\mathrm{e}$ obtained using the different interpolants and the exact neutronization rate obtained solving the NSE Saha equilibrium equations. The error goes to zero at the table nodes, but can reach up to 1, i.e. 100\% error, at low densities and high temperatures. The dashed lines show the most relevant combination of density and temperature for the SNIa $M_\mathrm{Ch}$ model in two conditions: when electron captures start in NSE and the neutronization rate is maximum (left columns) and when a value of $Y_\mathrm{e}=0.47$~mol~g$^{-1}$ is attained, a condition that is only reached in the innermost $\sim0.15$~\msunb of the WD (right columns). The WD mass shells go through $\rho$-$T$ conditions for which the relative error in $\dot{Y}_\mathrm{e}$ is as high as $5-20$~\% for the most simple interpolators: linlog, linlin, loglog and poly. On the other hand, the upper bound on the maximum error in the same conditions is $\sim3-5$~\% when either the spline or the pe$\nu$n interpolator is used. 

Table~\ref{t:2} shows the impact of the different interpolators on the final yields of the $M_\mathrm{Ch}$ model, under the heading ''NSE table interpolation``. The nucleosynthetic results confirm the intuition gained with Figs.~\ref{f:2} and \ref{f:3}: the most accurate interpolators are the cubic spline and pe$\nu$n, which perform almost equally well. All interpolants lead to accurate values of the final kinetic energy and just negligible errors in the mass of \nifsxb synthesized. The relevant differences appear when one looks into the nucleosynthesis of less abundant species. For instance, using linlog, loglog and poly leads to errors in the ejected mass of \nifsvb and \feffb of about 10\%, average errors in the isotopic yields of $20-30$\% (indicator $c_2$), and order-of-magnitude errors in the yields of some isotopes with yields smaller than $10^{-3}$~\msunb (indicators $e_2$ and $e_3$). On the other hand, the maximum isotopic error obtained with the cubic spline interpolator is 0.8~\% and that obtained with pe$\nu$n is 4.2~\%. The behaviour of interpolator linlin is intermediate between the two groups above.

Table~\ref{t:2} also shows the result of increasing the resolution of the NSE table by up to ten times more density nodes (''den$\times$10``), with $\Delta\log\rho=0.025$. With a table of this size, even using a linlog interpolant gives very good results, comparable to both the spline and the pe$\nu$n interpolators with the original table with $\Delta\log\rho=0.25$.

In the calculations reported in the following sections, I work with simplified nuclear networks and use a table to obtain the properties of NSE matter, with the aim of testing as faithfully as possible the strategies commonly used in many SNIa explosion models. In these tests, it is important that the treatment of NSE matter does not introduce additional errors, beside those attributable to simplified networks. For this purpose, I use the cubic spline interpolator and the NSE table with $\Delta\log\rho=0.25$.

\section{Simplified nuclear networks}\label{s:4}

\begin{table}
 \caption{Simplified nuclear networks.
 \label{t:3}}
 \centering
 \begin{tabular}{lcccc}
 \hline\hline
\noalign{\smallskip}
  & iso7 & $13\alpha$ & net21 & net23 \\
$Z$ & $A$ & $A$ & $A$ & $A$ \\
 \hline\hline
\noalign{\smallskip}
 n & - & - & 1 & 1 \\
 p & - & - & 1 & 1 \\
He & 4 & 4 & 4 & 4 \\
 C & 12 & 12 & 12 & 12 \\
 O & 16 & 16 & 16 & 16 \\
Ne & 20 & 20 & 20 & 20,22 \\
Mg & 24 & 24 & 24 & 24,25 \\
Si & 28 & 28 & 28 & 28 \\
 S & - & 32 & 32 & 32 \\
Ar & - & 36 & 36 & 36 \\
Ca & - & 40 & 40 & 40 \\
Ti & - & 44 & 44 & 44 \\
Cr & - & 48 & 48 & 48 \\
Fe & - & 52 & 52,53,54, & 52,53,54 \\
 &  &  & 55,56 & 55,56 \\
Co & - & - & 55,56 & 55,56 \\  
Ni & 56 & 56 & 56 & 56 \\
\hline\hline
 \end{tabular}
\end{table}

\begin{figure*}
   \includegraphics[width=\textwidth]{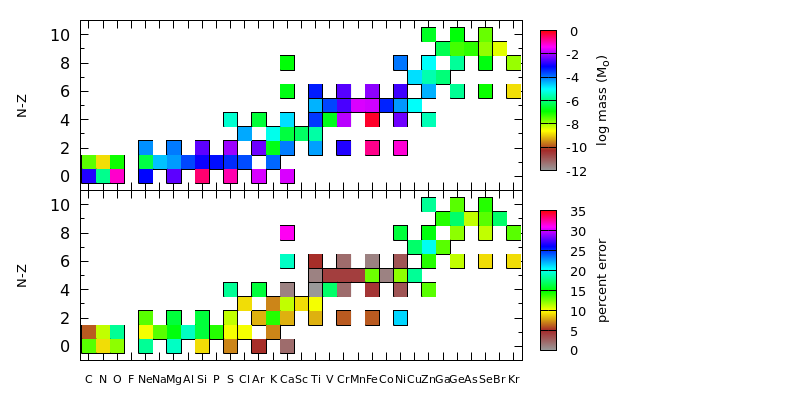}
\caption{Accuracy of the nucleosynthesis obtained with the net21 network as compared to model ddt2p4\_Z6p75e-2\_std. {\bf Top:} final ejected mass of stable isotopes in the reference model. {\bf Bottom:} percentage error of the mass yield of each isotope when the simplified network is used, with respect to the reference model.
}
\label{f:4}
\end{figure*}

Because of the huge difference between the size of a WD and the width of thermonuclear burning waves, multi-dimensional simulations of SNIa need to allocate most memory and CPU resources to solve the hydrodynamic equations over as large a range of length-scales as possible \citep[e.g.][]{2003gam}. In turn, the nuclear kinetics must be solved with a reduced nuclear network, with the goal that the nuclear energy must be released as faithfully as possible, as if a complete nuclear network were used. Other compositional properties that affect the equation of state, such as the electron mole number and the mean molar weight, must also be reproduced accurately in order not to change the explosion development. 
In this section, I show the impact of the use of reduced nuclear networks on the nucleosynthesis of both the $M_\mathrm{Ch}$ and the sub-$M_\mathrm{Ch}$ models in one dimension. 

First, I test two small networks, designed iso7 and $13\alpha$, that are widely used in multi-dimensional simulations of SNIa, plus a slightly larger network designed to improve the nucleosynthetic results, named net21. Table~\ref{t:3} shows the composition of each network. 
In these calculations, all nuclear reactions linking species present in the network directly are accounted for and the rates are taken from the REACLIB compilation (see Sect.~\ref{s:2}) unless otherwise stated here.

The iso7 network, introduced by \citet{2000tim} as a simplification of the nine-isotope reaction network described in detail in Table 1 of \citet{1986woo}, has been used in 2D simulations of DDT models of exploding WDs \citep[e.g.][]{2018leu}. This network is crafted for efficient computation of nuclear energy generation in multi-dimensional calculations of explosive burning stages from carbon-burning onwards. It assumes two quasi-equilibrium groups of isotopes, the silicon group and the iron group, and hard-wires the nucleosynthetic flows between both groups in a single step that link the abundances of \isotope{28}{Si} and \nifsxb.
This step is computed making use of Eqs.~6-8 of \citet{2000tim}.
\citet{2000tim} warned that the iso7 and $13\alpha$ networks (and, in general, any $\alpha$-network) might give energy generation rates wrong by order of magnitude if $Y_\mathrm{e}\lesssim0.49$~mol~g$^{-1}$. 

The $13\alpha$ network, introduced by \citet{1986mue} and later by \citet{1995liv} as a simplification of a larger network described in \citet{1978wea}, has been used in SPH simulations of merging and colliding WDs \citep{2014ras,2015dan}. The version of the $13\alpha$ network used in the present work follows \citet{1999tim} and \citet{2000tim}, where it is applied a special treatment to the links between $\alpha$-nuclei from magnesium onwards: above a temperature of $2.5\times10^9$~K, the flows from $(\alpha,\mathrm{p})$ reactions, followed by $(\mathrm{p},\gamma)$, are added to the flows from $(\alpha,\gamma)$. It is important to note that, in this version of the $13\alpha$ network\footnote{See also http://cococubed.asu.edu/code\_pages/burn\_helium.shtml}, the link from an $\alpha$-nucleus, \isotope{2Z}{Z}, to the next one, \isotope{2Z+4}{(Z+2)}, through $(\alpha,\mathrm{p})$ reactions takes into account the possibility that $(\mathrm{p},\alpha)$ follows instead of $(\mathrm{p},\gamma)$: 
\begin{equation}\label{e:0}
 ^{2Z}\mathrm{Z} \rightleftarrows ^{2Z+3}\mathrm{(Z+1)} \rightleftarrows ^{2Z+4}\mathrm{(Z+2)}\,.
\end{equation}
For instance, in the conversion of \isotope{28}{Si} into \isotope{32}{S} through \isotope{31}{P}, the four reactions that follow have to be considered besides $^{28}\mathrm{Si}(\alpha,\gamma)^{32}\mathrm{S}$,
\begin{equation}
 ^{28}\mathrm{Si} \rightleftarrows ^{31}\mathrm{P} \rightleftarrows ^{32}\mathrm{S}\,.
\end{equation}
Assuming that the abundance of the intermediate nucleus, \isotope{31}{P} in this example, is established by the equilibrium of the direct and reverse reactions in Eq.~\ref{e:0}, the overall rate of change of the molar fraction of \isotope{2Z+4}{(Z+2)}, $Y\left[^{2Z+4}\mathrm{(Z+2)}\right]$, from \isotope{2Z}{Z} is given by
\begin{equation}
 %\frac{\mathrm{d}Y\left[^{2Z+4}\mathrm{(Z+2)}\right]}{\mathrm{d}t} = 
 R_{(\alpha,\gamma)}^\mathrm{eff}
 Y\left[^{2Z}\mathrm{Z}\right] Y_{\alpha} - 
 R_{(\gamma,\alpha)}^\mathrm{eff}
 Y\left[^{2Z+4}\mathrm{(Z+2)}\right]\,,
\end{equation}
where the effective rates, $R_{(\alpha,\gamma)}^\mathrm{eff}$ and $R_{(\gamma,\alpha)}^\mathrm{eff}$, are
\begin{equation}\label{e:1}
 R_{(\alpha,\gamma)}^\mathrm{eff} = R_{(\alpha,\gamma)} + R_{(\alpha,\mathrm{p})}\frac{R_{(\mathrm{p},\gamma)}}{R_{(\mathrm{p},\alpha)} + R_{(\mathrm{p},\gamma)}}\,
\end{equation}
and
\begin{equation}\label{e:2}
 R_{(\gamma,\alpha)}^\mathrm{eff} = R_{(\gamma,\alpha)} + R_{(\gamma,\mathrm{p})}\frac{R_{(\mathrm{p},\alpha)}}{R_{(\mathrm{p},\alpha)} + R_{(\mathrm{p},\gamma)}}\,,
\end{equation}
and $R_{(\alpha,\mathrm{p})}$, $R_{(\mathrm{p},\alpha)}$, $R_{(\gamma,\mathrm{p})}$, and $R_{(\mathrm{p},\gamma)}$ are the true rates of the reactions \isotope{2Z}{Z}$\rightarrow$\isotope{2Z+3}{(Z+1)}, \isotope{2Z+3}{(Z+1)}$\rightarrow$\isotope{2Z}{Z}, \isotope{2Z+4}{(Z+2)}$\rightarrow$\isotope{2Z+3}{(Z+1)}, and \isotope{2Z+3}{(Z+1)}$\rightarrow$\isotope{2Z+4}{(Z+2)}, respectively. 
As before, all true rates have been computed as in the REACLIB compilation. 

In both networks, iso7 and $13\alpha$, the electron mole number in the initial model is $Y_\mathrm{e}=0.5$~\molg. The electron mole number is allowed to change during NSE, due to electron captures, but is kept fixed when the composition is computed by integration of the nuclear network, because it does not include weak interactions.

Table~\ref{t:2} shows the accuracy of the nucleosynthesis obtained by post-processing the thermodynamic trajectories belonging to these two networks, under the heading ''Simplified networks``. Generally, the performance of the simplified networks in the $M_\mathrm{ch}$ models is worse than in the sub-$M_\mathrm{Ch}$ models. The kinetic energy is reproduced reasonably with the simplified networks, slightly better with iso7 than with $13\alpha$, and the error in the yield of \isotope{56}{Ni} is within 6-10\% when iso7 is used but $\sim4\%$ with $13\alpha$.

The errors in the yields of radioisotopes \feffb and \nifsv, which are often constrained observationally by the late-time light curves of SNIa, are about 30\%, with the exception of \nifsvb in $M_\mathrm{Ch}$, the abundance of which is wrong by nearly an order of magnitude. These errors are comparable with the maximum deviation obtained in the yields of the most abundant elements, $d_1\sim 40 - 60\%$, while the maximum error in the abundance of the isotopes (indicators $e_1$ to $e_3$) may be up to several orders of magnitude. Among those with the largest yields, the isotopes that present the maximum deviation are mostly part of the iron group: \isotope{40}{Ca}, \isotope{52,53}{Cr}, \isotope{55}{Mn}, \isotope{57}{Fe}, and \isotope{60,62}{Ni}.

The net21 network is an extension of the $13\alpha$ network
that includes additional isotopes of the iron group plus free protons and neutrons (its full composition can be seen in Table~\ref{t:3}), in order to obtain a more reliable representation of the nucleosynthesis of the most deficient nuclides from the results of the iso7 and $13\alpha$ networks. 
The net21 network makes use of the same hard-wiring of rates as in Eqs.~\ref{e:1} and \ref{e:2}, with the exception of the chain
\begin{equation}
 ^{52}\mathrm{Fe} \rightleftarrows ^{55}\mathrm{Co} \rightleftarrows ^{56}\mathrm{Ni}\,.
\end{equation}
Since \isotope{55}{Co} is included explicitly in the network, the effective rates between \isotope{52}{Fe} and \isotope{56}{Ni} given by Eqs.~\ref{e:1} and \ref{e:2} are substituted by the corresponding true rates, $R_{(\alpha,\gamma)}$ and $R_{(\gamma,\alpha)}$. 

When using the net21 network in the hydrocode, the errors in the kinetic energy and the yield of \isotope{56}{Ni} after post-processing are about $1 - 5\%$. On the other hand, the errors in the yields of \feffb and \nifsvb are less than $7\%$ and the maximum error in the predicted abundance of any isotope or element is less than 20\% (indicators $d_1$ to $e_3$).

The errors in the post-processed nucleosynthesis after using the net21 network in the hydrocode are comparable to, although slightly larger than, those obtained with the large networks used in the convergence study, netAkh and nse7. Therefore, the performance of net21 would barely be improved with other simplified networks based on no more than a few tenths of nuclides. I have experimented with larger networks, including a group of CNO isotopes (\isotope{13,14}{C}, \isotope{14}{N}, \isotope{17}{O}) and the intermediate species in Eq.~\ref{e:0}, from \isotope{27}{Al} to \isotope{51}{Mn}, with no significant improvement in the performance over that of network net21. I have also probed reducing the number of iron-group isotopes in the network, but the accuracy of the nucleosynthesis was worse than with net21.

\section{Nuclear post-processing for high metallicity progenitors}\label{s:5}

In this section, I test the accuracy of the post-processed nucleosynthesis with respect to the initial metallicity of the progenitor star. Network net21, as well as iso7 and $13\alpha$, is not capable to describe an initial composition of carbon-oxygen material with an excess of neutrons over protons. This is because the initial metallicity of the progenitor star is encoded, at the time of formation of a carbon-oxygen WD, in the abundance of \isotope{22}{Ne} \citep{2003tim}, which is not a part of any of the simplified networks discussed in the previous section. Hence one may wonder whether the accuracy of the net21 network degrades for high metallicity progenitors.

Here, I introduce a new network, net23, that complements the net21 network with the inclusion of \isotope{22}{Ne} and \isotope{25}{Mg} (Table~\ref{t:3}). As just explained, the presence of \isotope{22}{Ne} serves the purpose of building initial models with non-zero neutron-excess, as it is done in the hydrodynamic models that use the full network (see Sect.~\ref{s:2}). The isotope \isotope{25}{Mg} provides a simple route for the burning of \isotope{22}{Ne} through \isotope{22}{Ne}$\left(\alpha,\mathrm{n}\right)$\isotope{25}{Mg}$\left(\alpha,\mathrm{n}\right)$\isotope{28}{Si}.

For this test, I have selected an initial metallicity of the WD as high as $Z=0.0675$. Table~\ref{t:2} shows the results of the post-processed nucleosynthesis using networks net21 and net23, for both the $M_\mathrm{Ch}$ and the sub-$M_\mathrm{Ch}$ models, under the heading ''High metallicity progenitor``. Network net21 performs slightly better in the $M_\mathrm{Ch}$ model, whereas net23 does better in the sub-$M_\mathrm{Ch}$ model, but the overall accuracy of both networks is similar. When the nucleosynthesis errors in the high metallicity calculations are compared with those in the $Z=0.009$ models, using the same networks, the results are slightly better at low metallicity, but not significantly different. 

Figure~\ref{f:4} shows the percentage error in the prediction of the abundances of the stable isotopes between carbon and krypton after post-processing the thermodynamic trajectories obtained with the net21 network, compared with those using the default network in the hydrocode. The largest error, 31\%, belongs to \isotope{48}{Ca}, the abundance of which is below $10^{-6}$~\msun, while all other isotopes are predicted with errors smaller than $\sim25\%$.

\section{Transition from the nuclear network to NSE and vice versa}

The simultaneous inclusion in a simulation of a simplified network and an NSE routine, as described in Sect.~\ref{s:3}, raises the question of the criteria for the transition between both treatments. The transition between the nuclear network and NSE is usually defined in terms of the temperature, where different values are used for assuming NSE, $T_\mathrm{NSE}$, and leaving it, $T_\mathrm{out}$. In the hydrocode used in this work, a third parameter, $\rho_\mathrm{NSE0}$, allows acceleration of the burning of shells hit by a deflagration front. 

Silicon exhaustion is a milestone for achieving NSE. It is reached at a temperature somewhere in between $\sim5\times10^9$~K and $\sim6\times10^9$~K with a slight dependence on density \citep[e.g., Fig.~20 in][]{1973woo}. In the calculations presented in this section, I have adopted a unified value of $T_\mathrm{NSE}$, independent of matter density\footnote{In the hydrocode using the default network, there are two different values of the minimum temperature to achieve NSE, depending on density; see \citet{2019bra} for further details}, as detailed in Table~\ref{t:2}.

The freezing-out of nuclear reactions when NSE matter cools is more complex, because at low densities the abundance of free particles (especially $\alpha$ particles) may be sufficiently large to affect the composition significantly. 
\citet{2017har} studied the impact of the value of $T_\mathrm{out}$ in the context of core-collapse supernovae.
Usual values of the temperature at which matter is assumed to leave NSE lie in the range from $\sim2\times10^9$~K to $\sim5\times10^9$~K. Again, in the calculations presented in this section, there is a single value of $T_\mathrm{out}$, independent of density, at variance with the method adopted in the hydrocode using the default network \citep{2019bra}.

Following the passage of a deflagrative front, shells are assumed to achieve NSE if their density is larger than the third parameter, $\rho_\mathrm{NSE0}$. By default, I adopt conservative values for the three parameters: $T_\mathrm{NSE}=6\times10^9$~K, $T_\mathrm{out}=5\times10^9$~K and $\rho_\mathrm{NSE0}=8\times10^7$~\gcc.

As can be seen in Table~\ref{t:2}, the precise value of $T_\mathrm{NSE}$ does not affect the error in the post-processed nucleosynthesis of the $M_\mathrm{Ch}$ model significantly, as long as it is in the range from $\sim5\times10^9$~K to $\sim6\times10^9$~K\footnote{Recall that the errors reported in this subsection have to be compared with the reference model, that is the net21 $M_\mathrm{Ch}$ model under the heading ''Simplified networks``.}.
On the other hand, when the threshold for leaving NSE, $T_\mathrm{out}$, takes on a value between $4\times10^9$~K and $5\times10^9$~K the accuracy of the nucleosynthesis is satisfactory, but when this parameter goes down to $3\times10^9$~K the results worsen: for instance, the maximum error in the predicted isotopic yields increases by orders of magnitude. 
Finally, the accuracy of the nucleosynthesis is not affected by the value of $\rho_\mathrm{NSE0}$, at least within the range explored in this work and presented in Table~\ref{t:2}, $\rho_\mathrm{NSE0}=4\times10^7$~\gccb to $10^8$~\gcc.

As explained in Appendix B4 of \citet{2019bra}, detonated matter is not assumed to be in NSE if its density is below $\rho_\mathrm{NSE0}$, irrespective of its temperature. In practice, it implies that, in the present models, freeze-out from NSE occurs at low entropy. Therefore, the effect of the NSE parameters just described limits to the so-called normal or particle-poor freeze-out. On the other hand, it is remarkable that alpha-rich freeze-out is managed by integration of the net21 network with very good accuracy, in spite of its small size.

\section{Conclusions}\label{s:conclusions}

Three-dimensional hydrodynamical simulations are necessary to predict the outcome of several explosion scenarios and compare them with SNIa observational data. Whereas great efforts have been made to obtain reliable nucleosynthetic yields and explore their dependence on a number of simulation parameters \citep[e.g.][]{2010mae,2013sei}, 
until now very few works have been published addressing the accuracy of the resulting nucleosynthesis with respect to the use of simplified nuclear networks \citep{2015pap}\footnote{https://trace.tennessee.edu/utk\_graddiss/3454/}.
In the present work, I use a supernova code, capable of integrating the nuclear kinetic equations using a large nuclear network and the hydrodynamical equations simultaneously, as a benchmark for testing the results of several simplifying assumptions related to nuclear kinetics. These simplifications are related to the use of a small nuclear network and the use of tabulated properties of matter in nuclear statistical equilibrium. I define a set of 14 indicators related to the accuracy of the nucleosynthesis.

The method used in this work does not allow testing of all the strategies currently used in multi-dimensional simulations of SNIa to follow the nuclear energy generation-rate accurately during a thermonuclear explosion. For instance, several studies \citep[e.g.][]{2004tra,2012dub,2017leu} use Lagrangian tracer particles smartly distributed through the simulated space to advect the thermodynamic properties of the underlying Eulerian cells. Nucleosynthesis is then obtained following a post-processing step on the tracer particles, the number of which is much less than the original Eulerian nodes of the simulation. Other studies \citep[e.g.][]{2007cal,2010fin,2016tow} adopt a nuclear energy generation rate linked to the nature of the burning wave (whether a detonation or a deflagration) and the density of fuel. 

I propose to use published tables of NSE properties (neutronization rate, mean molar number, mean nuclear binding energy and neutrino energy loss-rate, as functions of density, temperature and electron molar number)
but putting great care into interpolation between their values at the tabulated density points. The interpolation in temperature and electron mole number is not so critical and can simply be linear. A cubic spline interpolation in density gives the most precise results, the accuracy of which is better than 1\% in all 14 nucleosynthetic indicators. 
Alternatively, a simple linear interpolation on an NSE table with high resolution in density, $\Delta\log\rho=0.025$, may give almost as accurate results as the cubic spline interpolation of the standard NSE table with $\Delta\log\rho=0.25$.

I have tested several simplified nuclear networks for the accuracy of the nucleosynthesis obtained after post-processing, compared with the nucleosynthesis resulting directly from the supernova code when the default, large nuclear network is used. Short networks (iso7 and 13$\alpha$) are able to give an accurate yield of \nifsxb after post-processing, but can fail by an order of magnitude in predicting the ejected mass of even mildly abundant species ($>10^{-3}$~\msun). I find that a network of 21 species, net21, reproduces the nucleosynthesis of Chandrasekhar and sub-Chandrasekhar explosions nicely, their average errors being better than 10\% for the most abundant elements and isotopes (yields larger than $10^{-3}$~\msun) and better than 20\% for the whole set of stable elements and isotopes followed in the model.

In these explosion models, the NSE state is switched on and off according to three criteria based on local temperature and density. 
The temperature at which it can be safely assumed that matter will achieve NSE can adopt any value between $5\times10^9$~K and $6\times10^9$~K without affecting the accuracy of the post-processed nucleosynthesis significantly. For normal freeze-out of NSE, equilibrium abundances can be assumed for temperatures in excess of $\sim4\times10^9$~K. However, if the NSE state is kept until a temperature as low as $3\times10^9$~K, the resulting post-processed nucleosynthesis can be almost an order of magnitude wrong even for isotopes with yield $>10^{-3}$~\msun.
On the other hand, the alpha-rich freeze-out of NSE is managed satisfactorily by the net21 network, in spite of its small size.

\section*{Acknowledgements}

This work has been supported by the MINECO-FEDER grants AYA2015-63588-P and PGC2018-095317-B-C21.

\bibliographystyle{mnras} 
\bibliography{../ebg}

\begin{thebibliography}{}
\makeatletter
\relax
\def\mn@urlcharsother{\let\do\@makeother \do\$\do\&\do\#\do\^\do\_\do\%\do\~}
\def\mn@doi{\begingroup\mn@urlcharsother \@ifnextchar [ {\mn@doi@}
  {\mn@doi@[]}}
\def\mn@doi@[#1]#2{\def\@tempa{#1}\ifx\@tempa\@empty \href
  {http://dx.doi.org/#2} {doi:#2}\else \href {http://dx.doi.org/#2} {#1}\fi
  \endgroup}
\def\mn@eprint#1#2{\mn@eprint@#1:#2::\@nil}
\def\mn@eprint@arXiv#1{\href {http://arxiv.org/abs/#1} {{\tt arXiv:#1}}}
\def\mn@eprint@dblp#1{\href {http://dblp.uni-trier.de/rec/bibtex/#1.xml}
  {dblp:#1}}
\def\mn@eprint@#1:#2:#3:#4\@nil{\def\@tempa {#1}\def\@tempb {#2}\def\@tempc
  {#3}\ifx \@tempc \@empty \let \@tempc \@tempb \let \@tempb \@tempa \fi \ifx
  \@tempb \@empty \def\@tempb {arXiv}\fi \@ifundefined
  {mn@eprint@\@tempb}{\@tempb:\@tempc}{\expandafter \expandafter \csname
  mn@eprint@\@tempb\endcsname \expandafter{\@tempc}}}

\bibitem[\protect\citeauthoryear{{Arsenijevic}, {Fabbro}, {Mour{\~a}o}  \&
  {Rica da Silva}}{{Arsenijevic} et~al.}{2008}]{2008ars}
{Arsenijevic} V.,  {Fabbro} S.,  {Mour{\~a}o} A.~M.,   {Rica da Silva} A.~J.,
  2008, \mn@doi [\aap] {10.1051/0004-6361:200810675}, \href
  {https://ui.adsabs.harvard.edu/abs/2008A&A...492..535A} {492, 535}

\bibitem[\protect\citeauthoryear{{Badenes}, {Borkowski}, {Hughes}, {Hwang}  \&
  {Bravo}}{{Badenes} et~al.}{2006}]{2006bad}
{Badenes} C.,  {Borkowski} K.~J.,  {Hughes} J.~P.,  {Hwang} U.,   {Bravo} E.,
  2006, \mn@doi [\apj] {10.1086/504399}, \href
  {http://cdsads.u-strasbg.fr/abs/2006ApJ...645.1373B} {645, 1373}

\bibitem[\protect\citeauthoryear{{Blondin}, {Dessart}, {Hillier}  \&
  {Khokhlov}}{{Blondin} et~al.}{2013}]{2013blo}
{Blondin} S.,  {Dessart} L.,  {Hillier} D.~J.,   {Khokhlov} A.~M.,  2013,
  \mn@doi [\mnras] {10.1093/mnras/sts484}, \href
  {http://cdsads.u-strasbg.fr/abs/2013MNRAS.429.2127B} {429, 2127}

\bibitem[\protect\citeauthoryear{{Blondin}, {Dessart}, {Hillier}  \&
  {Khokhlov}}{{Blondin} et~al.}{2017}]{2017blo}
{Blondin} S.,  {Dessart} L.,  {Hillier} D.~J.,   {Khokhlov} A.~M.,  2017,
  \mn@doi [\mnras] {10.1093/mnras/stw2492}, \href
  {http://cdsads.u-strasbg.fr/abs/2017MNRAS.470..157B} {470, 157}

\bibitem[\protect\citeauthoryear{{Brachwitz} et~al.,}{{Brachwitz}
  et~al.}{2000}]{2000brc}
{Brachwitz} F.,  et~al., 2000, \mn@doi [\apj] {10.1086/308968}, \href
  {http://cdsads.u-strasbg.fr/abs/2000ApJ...536..934B} {536, 934}

\bibitem[\protect\citeauthoryear{{Branch}, {Fisher}  \& {Nugent}}{{Branch}
  et~al.}{1993}]{1993bran}
{Branch} D.,  {Fisher} A.,   {Nugent} P.,  1993, \mn@doi [\aj]
  {10.1086/116810}, \href
  {https://ui.adsabs.harvard.edu/abs/1993AJ....106.2383B} {106, 2383}

\bibitem[\protect\citeauthoryear{{Branch}, {Chau Dang}  \& {Baron}}{{Branch}
  et~al.}{2009}]{2009bran}
{Branch} D.,  {Chau Dang} L.,   {Baron} E.,  2009, \mn@doi [\pasp]
  {10.1086/597788}, \href
  {https://ui.adsabs.harvard.edu/abs/2009PASP..121..238B} {121, 238}

\bibitem[\protect\citeauthoryear{{Bravo}}{{Bravo}}{2019}]{2019brab}
{Bravo} E.,  2019, \mn@doi [\aap] {10.1051/0004-6361/201935095}, \href
  {https://ui.adsabs.harvard.edu/abs/2019A%26A...624A.139B} {624, A139}

\bibitem[\protect\citeauthoryear{{Bravo} \& {Garc{\'i}a-Senz}}{{Bravo} \&
  {Garc{\'i}a-Senz}}{2006}]{2006bra}
{Bravo} E.,  {Garc{\'i}a-Senz} D.,  2006, \mn@doi [\apjl] {10.1086/504713},
  \href {http://cdsads.u-strasbg.fr/abs/2006ApJ...642L.157B} {642, L157}

\bibitem[\protect\citeauthoryear{{Bravo} \& {Mart{\'i}nez-Pinedo}}{{Bravo} \&
  {Mart{\'i}nez-Pinedo}}{2012}]{2012bra}
{Bravo} E.,  {Mart{\'i}nez-Pinedo} G.,  2012, \mn@doi [\prc]
  {10.1103/PhysRevC.85.055805}, \href
  {http://cdsads.u-strasbg.fr/abs/2012PhRvC..85e5805B} {85, 055805}

\bibitem[\protect\citeauthoryear{{Bravo}, {Dom{\'i}nguez}, {Badenes},
  {Piersanti}  \& {Straniero}}{{Bravo} et~al.}{2010}]{2010bra}
{Bravo} E.,  {Dom{\'i}nguez} I.,  {Badenes} C.,  {Piersanti} L.,   {Straniero}
  O.,  2010, \mn@doi [\apjl] {10.1088/2041-8205/711/2/L66}, \href
  {http://cdsads.u-strasbg.fr/abs/2010ApJ...711L..66B} {711, L66}

\bibitem[\protect\citeauthoryear{{Bravo}, {Badenes}  \&
  {Mart{\'{\i}}nez-Rodr{\'{\i}}guez}}{{Bravo} et~al.}{2019}]{2019bra}
{Bravo} E.,  {Badenes} C.,   {Mart{\'{\i}}nez-Rodr{\'{\i}}guez} H.,  2019,
  \mn@doi [\mnras] {10.1093/mnras/sty2951}, \href
  {http://cdsads.u-strasbg.fr/abs/2019MNRAS.482.4346B} {482, 4346}

\bibitem[\protect\citeauthoryear{{Calder} et~al.,}{{Calder}
  et~al.}{2007}]{2007cal}
{Calder} A.~C.,  et~al., 2007, \mn@doi [\apj] {10.1086/510709}, \href
  {http://cdsads.u-strasbg.fr/abs/2007ApJ...656..313C} {656, 313}

\bibitem[\protect\citeauthoryear{{Churazov} et~al.,}{{Churazov}
  et~al.}{2014}]{2014chu}
{Churazov} E.,  et~al., 2014, \mn@doi [\nat] {10.1038/nature13672}, \href
  {http://cdsads.u-strasbg.fr/abs/2014Natur.512..406C} {512, 406}

\bibitem[\protect\citeauthoryear{{Churazov} et~al.,}{{Churazov}
  et~al.}{2015}]{2015chu}
{Churazov} E.,  et~al., 2015, \mn@doi [\apj] {10.1088/0004-637X/812/1/62},
  \href {http://cdsads.u-strasbg.fr/abs/2015ApJ...812...62C} {812, 62}

\bibitem[\protect\citeauthoryear{{Cyburt} et~al.,}{{Cyburt}
  et~al.}{2010}]{2010cyb}
{Cyburt} R.~H.,  et~al., 2010, \mn@doi [\apjs] {10.1088/0067-0049/189/1/240},
  \href {http://cdsads.u-strasbg.fr/abs/2010ApJS..189..240C} {189, 240}

\bibitem[\protect\citeauthoryear{{Dan}, {Guillochon}, {Br{\"u}ggen},
  {Ramirez-Ruiz}  \& {Rosswog}}{{Dan} et~al.}{2015}]{2015dan}
{Dan} M.,  {Guillochon} J.,  {Br{\"u}ggen} M.,  {Ramirez-Ruiz} E.,   {Rosswog}
  S.,  2015, \mn@doi [\mnras] {10.1093/mnras/stv2289}, \href
  {https://ui.adsabs.harvard.edu/abs/2015MNRAS.454.4411D} {454, 4411}

\bibitem[\protect\citeauthoryear{{Diehl} et~al.,}{{Diehl}
  et~al.}{2014}]{2014die}
{Diehl} R.,  et~al., 2014, \mn@doi [Science] {10.1126/science.1254738}, \href
  {http://cdsads.u-strasbg.fr/abs/2014Sci...345.1162D} {345, 1162}

\bibitem[\protect\citeauthoryear{{Dimitriadis} et~al.,}{{Dimitriadis}
  et~al.}{2017}]{2017dim}
{Dimitriadis} G.,  et~al., 2017, \mn@doi [\mnras] {10.1093/mnras/stx683}, \href
  {http://cdsads.u-strasbg.fr/abs/2017MNRAS.468.3798D} {468, 3798}

\bibitem[\protect\citeauthoryear{{Dubey}, {Daley}, {ZuHone}, {Ricker}, {Weide}
  \& {Graziani}}{{Dubey} et~al.}{2012}]{2012dub}
{Dubey} A.,  {Daley} C.,  {ZuHone} J.,  {Ricker} P.~M.,  {Weide} K.,
  {Graziani} C.,  2012, \mn@doi [\apjs] {10.1088/0067-0049/201/2/27}, \href
  {https://ui.adsabs.harvard.edu/abs/2012ApJS..201...27D} {201, 27}

\bibitem[\protect\citeauthoryear{{Filippenko}}{{Filippenko}}{1997}]{1997fil}
{Filippenko} A.~V.,  1997, \mn@doi [\araa] {10.1146/annurev.astro.35.1.309},
  \href {https://ui.adsabs.harvard.edu/abs/1997ARA&A..35..309F} {35, 309}

\bibitem[\protect\citeauthoryear{{Fink}, {R{\"o}pke}, {Hillebrandt},
  {Seitenzahl}, {Sim}  \& {Kromer}}{{Fink} et~al.}{2010}]{2010fin}
{Fink} M.,  {R{\"o}pke} F.~K.,  {Hillebrandt} W.,  {Seitenzahl} I.~R.,  {Sim}
  S.~A.,   {Kromer} M.,  2010, \mn@doi [\aap] {10.1051/0004-6361/200913892},
  \href {http://cdsads.u-strasbg.fr/abs/2010A\%26A...514A..53F} {514, A53+}

\bibitem[\protect\citeauthoryear{{Fuller}, {Fowler}  \& {Newman}}{{Fuller}
  et~al.}{1985}]{1985ful}
{Fuller} G.~M.,  {Fowler} W.~A.,   {Newman} M.~J.,  1985, \mn@doi [\apj]
  {10.1086/163208}, \href {http://adsabs.harvard.edu/abs/1985ApJ...293....1F}
  {293, 1}

\bibitem[\protect\citeauthoryear{{Gamezo}, {Khokhlov}, {Oran}, {Chtchelkanova}
  \& {Rosenberg}}{{Gamezo} et~al.}{2003}]{2003gam}
{Gamezo} V.~N.,  {Khokhlov} A.~M.,  {Oran} E.~S.,  {Chtchelkanova} A.~Y.,
  {Rosenberg} R.~O.,  2003, \mn@doi [Science] {10.1126/science.1078129}, \href
  {http://cdsads.u-strasbg.fr/abs/2003Sci...299...77G} {299, 77}

\bibitem[\protect\citeauthoryear{{Graur}, {Zurek}, {Shara}, {Riess},
  {Seitenzahl}  \& {Rest}}{{Graur} et~al.}{2016}]{2016gra}
{Graur} O.,  {Zurek} D.,  {Shara} M.~M.,  {Riess} A.~G.,  {Seitenzahl} I.~R.,
  {Rest} A.,  2016, \mn@doi [\apj] {10.3847/0004-637X/819/1/31}, \href
  {https://ui.adsabs.harvard.edu/abs/2016ApJ...819...31G} {819, 31}

\bibitem[\protect\citeauthoryear{{Graur} et~al.,}{{Graur}
  et~al.}{2018}]{2018gra}
{Graur} O.,  et~al., 2018, \mn@doi [\apj] {10.3847/1538-4357/aabe25}, \href
  {http://cdsads.u-strasbg.fr/abs/2018ApJ...859...79G} {859, 79}

\bibitem[\protect\citeauthoryear{{Harris}, {Hix}, {Chertkow}, {Lee}, {Lentz}
  \& {Messer}}{{Harris} et~al.}{2017}]{2017har}
{Harris} J.~A.,  {Hix} W.~R.,  {Chertkow} M.~A.,  {Lee} C.~T.,  {Lentz} E.~J.,
   {Messer} O.~E.~B.,  2017, \mn@doi [\apj] {10.3847/1538-4357/aa76de}, \href
  {https://ui.adsabs.harvard.edu/abs/2017ApJ...843....2H} {843, 2}

\bibitem[\protect\citeauthoryear{{Hillebrandt} \& {Niemeyer}}{{Hillebrandt} \&
  {Niemeyer}}{2000}]{2000hil}
{Hillebrandt} W.,  {Niemeyer} J.~C.,  2000, \mn@doi [Annual Review of Astronomy
  and Astrophysics] {10.1146/annurev.astro.38.1.191}, \href
  {http://cdsads.u-strasbg.fr/abs/2000ARA\%26A..38..191H} {38, 191}

\bibitem[\protect\citeauthoryear{{Hoeflich} et~al.,}{{Hoeflich}
  et~al.}{2017}]{2017hoe}
{Hoeflich} P.,  et~al., 2017, preprint, \href
  {http://cdsads.u-strasbg.fr/abs/2017arXiv170705350H} {} (\mn@eprint {arXiv}
  {1707.05350})

\bibitem[\protect\citeauthoryear{{H{\"o}flich} \& {Khokhlov}}{{H{\"o}flich} \&
  {Khokhlov}}{1996}]{1996hoe}
{H{\"o}flich} P.,  {Khokhlov} A.,  1996, \mn@doi [\apj] {10.1086/176748}, \href
  {http://cdsads.u-strasbg.fr/abs/1996ApJ...457..500H} {457, 500}

\bibitem[\protect\citeauthoryear{{Hosseinzadeh} et~al.,}{{Hosseinzadeh}
  et~al.}{2017}]{2017hos}
{Hosseinzadeh} G.,  et~al., 2017, \mn@doi [\apj] {10.3847/2041-8213/aa8402},
  \href {https://ui.adsabs.harvard.edu/abs/2017ApJ...845L..11H} {845, L11}

\bibitem[\protect\citeauthoryear{{Howell}}{{Howell}}{2011}]{2011how}
{Howell} D.~A.,  2011, \mn@doi [Nature Communications] {10.1038/ncomms1344},
  \href {http://cdsads.u-strasbg.fr/abs/2011NatCo...2E.350H} {2, 350}

\bibitem[\protect\citeauthoryear{{Howell}, {H{\"o}flich}, {Wang}  \&
  {Wheeler}}{{Howell} et~al.}{2001}]{2001how}
{Howell} D.~A.,  {H{\"o}flich} P.,  {Wang} L.,   {Wheeler} J.~C.,  2001,
  \mn@doi [\apj] {10.1086/321584}, \href
  {http://adsabs.harvard.edu/abs/2001ApJ...556..302H} {556, 302}

\bibitem[\protect\citeauthoryear{{Howell} et~al.,}{{Howell}
  et~al.}{2005}]{2005how}
{Howell} D.~A.,  et~al., 2005, \mn@doi [\apj] {10.1086/497119}, \href
  {https://ui.adsabs.harvard.edu/abs/2005ApJ...634.1190H} {634, 1190}

\bibitem[\protect\citeauthoryear{{Isern} et~al.,}{{Isern}
  et~al.}{2016}]{2016ise}
{Isern} J.,  et~al., 2016, \mn@doi [\aap] {10.1051/0004-6361/201526941}, \href
  {http://cdsads.u-strasbg.fr/abs/2016A\%26A...588A..67I} {588, A67}

\bibitem[\protect\citeauthoryear{{Jacobson-Gal{\'a}n}, {Dimitriadis}, {Foley}
  \& {Kilpatrick}}{{Jacobson-Gal{\'a}n} et~al.}{2018}]{2018jac}
{Jacobson-Gal{\'a}n} W.~V.,  {Dimitriadis} G.,  {Foley} R.~J.,   {Kilpatrick}
  C.~D.,  2018, \mn@doi [\apj] {10.3847/1538-4357/aab716}, \href
  {http://cdsads.u-strasbg.fr/abs/2018ApJ...857...88J} {857, 88}

\bibitem[\protect\citeauthoryear{{James}, {Davis}, {Schmidt}  \& {Kim}}{{James}
  et~al.}{2006}]{2006jam}
{James} J.~B.,  {Davis} T.~M.,  {Schmidt} B.~P.,   {Kim} A.~G.,  2006, \mn@doi
  [\mnras] {10.1111/j.1365-2966.2006.10508.x}, \href
  {https://ui.adsabs.harvard.edu/abs/2006MNRAS.370..933J} {370, 933}

\bibitem[\protect\citeauthoryear{{Kasen} \& {Woosley}}{{Kasen} \&
  {Woosley}}{2007}]{2007kasb}
{Kasen} D.,  {Woosley} S.~E.,  2007, \mn@doi [\apj] {10.1086/510375}, \href
  {http://cdsads.u-strasbg.fr/abs/2007ApJ...656..661K} {656, 661}

\bibitem[\protect\citeauthoryear{{Katz} \& {Dong}}{{Katz} \&
  {Dong}}{2012}]{2012kaz}
{Katz} B.,  {Dong} S.,  2012, arXiv e-prints, \href
  {https://ui.adsabs.harvard.edu/abs/2012arXiv1211.4584K} {p. arXiv:1211.4584}

\bibitem[\protect\citeauthoryear{{Khokhlov}}{{Khokhlov}}{1991}]{1991kho}
{Khokhlov} A.~M.,  1991, \aap, \href
  {http://cdsads.u-strasbg.fr/abs/1991A\%26A...245..114K} {245, 114}

\bibitem[\protect\citeauthoryear{{Kushnir}}{{Kushnir}}{2019}]{2019kus}
{Kushnir} D.,  2019, \mn@doi [\mnras] {10.1093/mnras/sty3121}, \href
  {https://ui.adsabs.harvard.edu/abs/2019MNRAS.483..425K} {483, 425}

\bibitem[\protect\citeauthoryear{{Kushnir} \& {Katz}}{{Kushnir} \&
  {Katz}}{2019}]{2019kusb}
{Kushnir} D.,  {Katz} B.,  2019, arXiv e-prints, \href
  {https://ui.adsabs.harvard.edu/abs/2019arXiv191206151K} {p. arXiv:1912.06151}

\bibitem[\protect\citeauthoryear{{Kushnir}, {Katz}, {Dong}, {Livne}  \&
  {Fern{\'a}ndez}}{{Kushnir} et~al.}{2013}]{2013kus}
{Kushnir} D.,  {Katz} B.,  {Dong} S.,  {Livne} E.,   {Fern{\'a}ndez} R.,  2013,
  \mn@doi [\apj] {10.1088/2041-8205/778/2/L37}, \href
  {https://ui.adsabs.harvard.edu/abs/2013ApJ...778L..37K} {778, L37}

\bibitem[\protect\citeauthoryear{{Leung} \& {Nomoto}}{{Leung} \&
  {Nomoto}}{2017}]{2017leu}
{Leung} S.-C.,  {Nomoto} K.,  2017, preprint, \href
  {http://cdsads.u-strasbg.fr/abs/2017arXiv171004254L} {} (\mn@eprint {arXiv}
  {1710.04254})

\bibitem[\protect\citeauthoryear{{Leung} \& {Nomoto}}{{Leung} \&
  {Nomoto}}{2018}]{2018leu}
{Leung} S.-C.,  {Nomoto} K.,  2018, \mn@doi [\apj] {10.3847/1538-4357/aac2df},
  \href {https://ui.adsabs.harvard.edu/abs/2018ApJ...861..143L} {861, 143}

\bibitem[\protect\citeauthoryear{{Li} et~al.,}{{Li} et~al.}{2019}]{2019li}
{Li} W.,  et~al., 2019, arXiv e-prints, \href
  {https://ui.adsabs.harvard.edu/abs/2019arXiv190607321L} {p. arXiv:1906.07321}

\bibitem[\protect\citeauthoryear{{Livne} \& {Arnett}}{{Livne} \&
  {Arnett}}{1995}]{1995liv}
{Livne} E.,  {Arnett} D.,  1995, \mn@doi [\apj] {10.1086/176279}, \href
  {https://ui.adsabs.harvard.edu/abs/1995ApJ...452...62L} {452, 62}

\bibitem[\protect\citeauthoryear{{Lopez}, {Ramirez-Ruiz}, {Huppenkothen},
  {Badenes}  \& {Pooley}}{{Lopez} et~al.}{2011}]{2011lop}
{Lopez} L.~A.,  {Ramirez-Ruiz} E.,  {Huppenkothen} D.,  {Badenes} C.,
  {Pooley} D.~A.,  2011, \mn@doi [\apj] {10.1088/0004-637X/732/2/114}, \href
  {http://cdsads.u-strasbg.fr/abs/2011ApJ...732..114L} {732, 114}

\bibitem[\protect\citeauthoryear{{Maeda} et~al.,}{{Maeda}
  et~al.}{2010a}]{2010maeb}
{Maeda} K.,  et~al., 2010a, \mn@doi [\nat] {10.1038/nature09122}, \href
  {http://cdsads.u-strasbg.fr/abs/2010Natur.466...82M} {466, 82}

\bibitem[\protect\citeauthoryear{{Maeda}, {R{\"o}pke}, {Fink}, {Hillebrandt},
  {Travaglio}  \& {Thielemann}}{{Maeda} et~al.}{2010b}]{2010mae}
{Maeda} K.,  {R{\"o}pke} F.~K.,  {Fink} M.,  {Hillebrandt} W.,  {Travaglio} C.,
    {Thielemann} F.,  2010b, \mn@doi [\apj] {10.1088/0004-637X/712/1/624},
  \href {http://cdsads.u-strasbg.fr/abs/2010ApJ...712..624M} {712, 624}

\bibitem[\protect\citeauthoryear{{Maguire} et~al.,}{{Maguire}
  et~al.}{2018}]{2018mag}
{Maguire} K.,  et~al., 2018, \mn@doi [\mnras] {10.1093/mnras/sty820}, \href
  {http://cdsads.u-strasbg.fr/abs/2018MNRAS.477.3567M} {477, 3567}

\bibitem[\protect\citeauthoryear{Maoz, Mannucci  \& Nelemans}{Maoz
  et~al.}{2014}]{2014mao}
Maoz D.,  Mannucci F.,   Nelemans G.,  2014, \mn@doi [Annual Review of
  Astronomy and Astrophysics] {10.1146/annurev-astro-082812-141031}, 52, 107

\bibitem[\protect\citeauthoryear{{Mart{\'{\i}}nez-Rodr{\'{\i}}guez}
  et~al.,}{{Mart{\'{\i}}nez-Rodr{\'{\i}}guez} et~al.}{2018}]{2018mar}
{Mart{\'{\i}}nez-Rodr{\'{\i}}guez} H.,  et~al., 2018, \mn@doi [\apj]
  {10.3847/1538-4357/aadaec}, \href
  {http://cdsads.u-strasbg.fr/abs/2018ApJ...865..151M} {865, 151}

\bibitem[\protect\citeauthoryear{{Maund} et~al.,}{{Maund}
  et~al.}{2010}]{2010mau}
{Maund} J.~R.,  et~al., 2010, \mn@doi [\apjl] {10.1088/2041-8205/725/2/L167},
  \href {https://ui.adsabs.harvard.edu/abs/2010ApJ...725L.167M} {725, L167}

\bibitem[\protect\citeauthoryear{{Maund} et~al.,}{{Maund}
  et~al.}{2013}]{2013mau}
{Maund} J.~R.,  et~al., 2013, \mn@doi [\mnras] {10.1093/mnrasl/slt050}, \href
  {https://ui.adsabs.harvard.edu/abs/2013MNRAS.433L..20M} {433, L20}

\bibitem[\protect\citeauthoryear{{Miller} et~al.,}{{Miller}
  et~al.}{2018}]{2018mie}
{Miller} A.~A.,  et~al., 2018, \mn@doi [\apj] {10.3847/1538-4357/aaa01f}, \href
  {https://ui.adsabs.harvard.edu/abs/2018ApJ...852..100M} {852, 100}

\bibitem[\protect\citeauthoryear{{Mueller}}{{Mueller}}{1986}]{1986mue}
{Mueller} E.,  1986, \aap, \href
  {https://ui.adsabs.harvard.edu/abs/1986A&A...162..103M} {162, 103}

\bibitem[\protect\citeauthoryear{{Nugent}, {Baron}, {Branch}, {Fisher}  \&
  {Hauschildt}}{{Nugent} et~al.}{1997}]{1997nug}
{Nugent} P.,  {Baron} E.,  {Branch} D.,  {Fisher} A.,   {Hauschildt} P.~H.,
  1997, \mn@doi [\apj] {10.1086/304459}, \href
  {http://cdsads.u-strasbg.fr/abs/1997ApJ...485..812N} {485, 812}

\bibitem[\protect\citeauthoryear{{Papatheodore}}{{Papatheodore}}{2015}]{2015pap}
{Papatheodore} T.~L.,  2015, PhD thesis, University of Tennessee, Knoxville

\bibitem[\protect\citeauthoryear{{Parikh}, {Jos{\'e}}, {Seitenzahl}  \&
  {R{\"o}pke}}{{Parikh} et~al.}{2013}]{2013pkh}
{Parikh} A.,  {Jos{\'e}} J.,  {Seitenzahl} I.~R.,   {R{\"o}pke} F.~K.,  2013,
  \mn@doi [\aap] {10.1051/0004-6361/201321518}, \href
  {http://cdsads.u-strasbg.fr/abs/2013A%26A...557A...3P} {557, A3}

\bibitem[\protect\citeauthoryear{{Plewa}, {Calder}  \& {Lamb}}{{Plewa}
  et~al.}{2004}]{2004ple}
{Plewa} T.,  {Calder} A.~C.,   {Lamb} D.~Q.,  2004, \mn@doi [\apjl]
  {10.1086/424036}, \href {http://cdsads.u-strasbg.fr/abs/2004ApJ...612L..37P}
  {612, L37}

\bibitem[\protect\citeauthoryear{{Poludnenko}, {Chambers}, {Ahmed}, {Gamezo}
  \& {Taylor}}{{Poludnenko} et~al.}{2019}]{2019pol}
{Poludnenko} A.~Y.,  {Chambers} J.,  {Ahmed} K.,  {Gamezo} V.~N.,   {Taylor}
  B.~D.,  2019, \mn@doi [Science] {10.1126/science.aau7365}, \href
  {https://ui.adsabs.harvard.edu/abs/2019Sci...366.7365P} {366, aau7365}

\bibitem[\protect\citeauthoryear{{Poznanski}, {Gal-Yam}, {Maoz}, {Filippenko},
  {Leonard}  \& {Matheson}}{{Poznanski} et~al.}{2002}]{2002poz}
{Poznanski} D.,  {Gal-Yam} A.,  {Maoz} D.,  {Filippenko} A.~V.,  {Leonard}
  D.~C.,   {Matheson} T.,  2002, \mn@doi [\pasp] {10.1086/341741}, \href
  {https://ui.adsabs.harvard.edu/abs/2002PASP..114..833P} {114, 833}

\bibitem[\protect\citeauthoryear{{Raskin}, {Kasen}, {Moll}, {Schwab}  \&
  {Woosley}}{{Raskin} et~al.}{2014}]{2014ras}
{Raskin} C.,  {Kasen} D.,  {Moll} R.,  {Schwab} J.,   {Woosley} S.,  2014,
  \mn@doi [\apj] {10.1088/0004-637X/788/1/75}, \href
  {http://cdsads.u-strasbg.fr/abs/2014ApJ...788...75R} {788, 75}

\bibitem[\protect\citeauthoryear{{R{\"o}pke}, {Hillebrandt}, {Niemeyer}  \&
  {Woosley}}{{R{\"o}pke} et~al.}{2006}]{2006rop}
{R{\"o}pke} F.~K.,  {Hillebrandt} W.,  {Niemeyer} J.~C.,   {Woosley} S.~E.,
  2006, \mn@doi [\aap] {10.1051/0004-6361:20053926}, \href
  {http://cdsads.u-strasbg.fr/abs/2006A\%26A...448....1R} {448, 1}

\bibitem[\protect\citeauthoryear{{Scalzo} et~al.,}{{Scalzo}
  et~al.}{2014}]{2014sca}
{Scalzo} R.,  et~al., 2014, \mn@doi [\mnras] {10.1093/mnras/stu350}, \href
  {http://cdsads.u-strasbg.fr/abs/2014MNRAS.440.1498S} {440, 1498}

\bibitem[\protect\citeauthoryear{{Seitenzahl}, {Townsley}, {Peng}  \&
  {Truran}}{{Seitenzahl} et~al.}{2009}]{2009se2}
{Seitenzahl} I.~R.,  {Townsley} D.~M.,  {Peng} F.,   {Truran} J.~W.,  2009,
  \mn@doi [Atomic Data and Nuclear Data Tables] {10.1016/j.adt.2008.08.001},
  \href {http://cdsads.u-strasbg.fr/abs/2009ADNDT..95...96S} {95, 96}

\bibitem[\protect\citeauthoryear{{Seitenzahl} et~al.,}{{Seitenzahl}
  et~al.}{2013}]{2013sei}
{Seitenzahl} I.~R.,  et~al., 2013, \mn@doi [\mnras] {10.1093/mnras/sts402},
  \href {https://ui.adsabs.harvard.edu/abs/2013MNRAS.429.1156S} {429, 1156}

\bibitem[\protect\citeauthoryear{{Shappee}, {Stanek}, {Kochanek}  \&
  {Garnavich}}{{Shappee} et~al.}{2017}]{2017sha}
{Shappee} B.~J.,  {Stanek} K.~Z.,  {Kochanek} C.~S.,   {Garnavich} P.~M.,
  2017, \mn@doi [\apj] {10.3847/1538-4357/aa6eab}, \href
  {https://ui.adsabs.harvard.edu/abs/2017ApJ...841...48S} {841, 48}

\bibitem[\protect\citeauthoryear{{Shen}, {Kasen}, {Miles}  \&
  {Townsley}}{{Shen} et~al.}{2018}]{2018she}
{Shen} K.~J.,  {Kasen} D.,  {Miles} B.~J.,   {Townsley} D.~M.,  2018, \mn@doi
  [\apj] {10.3847/1538-4357/aaa8de}, \href
  {http://cdsads.u-strasbg.fr/abs/2018ApJ...854...52S} {854, 52}

\bibitem[\protect\citeauthoryear{{Sim}, {R{\"o}pke}, {Hillebrandt}, {Kromer},
  {Pakmor}, {Fink}, {Ruiter}  \& {Seitenzahl}}{{Sim} et~al.}{2010}]{2010sim}
{Sim} S.~A.,  {R{\"o}pke} F.~K.,  {Hillebrandt} W.,  {Kromer} M.,  {Pakmor} R.,
   {Fink} M.,  {Ruiter} A.~J.,   {Seitenzahl} I.~R.,  2010, \mn@doi [\apjl]
  {10.1088/2041-8205/714/1/L52}, \href
  {http://adsabs.harvard.edu/abs/2010ApJ...714L..52S} {714, L52}

\bibitem[\protect\citeauthoryear{{Stritzinger}, {Mazzali}, {Sollerman}  \&
  {Benetti}}{{Stritzinger} et~al.}{2006}]{2006stt}
{Stritzinger} M.,  {Mazzali} P.~A.,  {Sollerman} J.,   {Benetti} S.,  2006,
  \mn@doi [\aap] {10.1051/0004-6361:20065514}, \href
  {http://cdsads.u-strasbg.fr/abs/2006A\%26A...460..793S} {460, 793}

\bibitem[\protect\citeauthoryear{{Tanaka}, {Mazzali}, {Stanishev}, {Maurer},
  {Kerzendorf}  \& {Nomoto}}{{Tanaka} et~al.}{2011}]{2011tan}
{Tanaka} M.,  {Mazzali} P.~A.,  {Stanishev} V.,  {Maurer} I.,  {Kerzendorf}
  W.~E.,   {Nomoto} K.,  2011, \mn@doi [\mnras]
  {10.1111/j.1365-2966.2010.17556.x}, \href
  {http://cdsads.u-strasbg.fr/abs/2011MNRAS.410.1725T} {410, 1725}

\bibitem[\protect\citeauthoryear{{Thielemann}, {Nomoto}  \&
  {Yokoi}}{{Thielemann} et~al.}{1986}]{1986thi}
{Thielemann} F.-K.,  {Nomoto} K.,   {Yokoi} K.,  1986, \aap, \href
  {http://cdsads.u-strasbg.fr/abs/1986A\%26A...158...17T} {158, 17}

\bibitem[\protect\citeauthoryear{{Timmes}}{{Timmes}}{1999}]{1999tim}
{Timmes} F.~X.,  1999, \mn@doi [\apjs] {10.1086/313257}, \href
  {https://ui.adsabs.harvard.edu/abs/1999ApJS..124..241T} {124, 241}

\bibitem[\protect\citeauthoryear{{Timmes}, {Hoffman}  \& {Woosley}}{{Timmes}
  et~al.}{2000}]{2000tim}
{Timmes} F.~X.,  {Hoffman} R.~D.,   {Woosley} S.~E.,  2000, \mn@doi [\apjs]
  {10.1086/313407}, \href {http://adsabs.harvard.edu/abs/2000ApJS..129..377T}
  {129, 377}

\bibitem[\protect\citeauthoryear{{Timmes}, {Brown}  \& {Truran}}{{Timmes}
  et~al.}{2003}]{2003tim}
{Timmes} F.~X.,  {Brown} E.~F.,   {Truran} J.~W.,  2003, \mn@doi [\apjl]
  {10.1086/376721}, \href {http://cdsads.u-strasbg.fr/abs/2003ApJ...590L..83T}
  {590, L83}

\bibitem[\protect\citeauthoryear{{Townsley}, {Miles}, {Timmes}, {Calder}  \&
  {Brown}}{{Townsley} et~al.}{2016}]{2016tow}
{Townsley} D.~M.,  {Miles} B.~J.,  {Timmes} F.~X.,  {Calder} A.~C.,   {Brown}
  E.~F.,  2016, \mn@doi [\apjs] {10.3847/0067-0049/225/1/3}, \href
  {http://cdsads.u-strasbg.fr/abs/2016ApJS..225....3T} {225, 3}

\bibitem[\protect\citeauthoryear{{Travaglio}, {Hillebrandt}, {Reinecke}  \&
  {Thielemann}}{{Travaglio} et~al.}{2004}]{2004tra}
{Travaglio} C.,  {Hillebrandt} W.,  {Reinecke} M.,   {Thielemann} F.~K.,  2004,
  \mn@doi [\aap] {10.1051/0004-6361:20041108}, \href
  {https://ui.adsabs.harvard.edu/abs/2004A&A...425.1029T} {425, 1029}

\bibitem[\protect\citeauthoryear{{Wang}, {Wheeler}, {Li}  \&
  {Clocchiatti}}{{Wang} et~al.}{1996}]{1996wan}
{Wang} L.,  {Wheeler} J.~C.,  {Li} Z.,   {Clocchiatti} A.,  1996, \mn@doi
  [\apj] {10.1086/177617}, \href
  {https://ui.adsabs.harvard.edu/abs/1996ApJ...467..435W} {467, 435}

\bibitem[\protect\citeauthoryear{{Weaver}, {Zimmerman}  \& {Woosley}}{{Weaver}
  et~al.}{1978}]{1978wea}
{Weaver} T.~A.,  {Zimmerman} G.~B.,   {Woosley} S.~E.,  1978, \mn@doi [\apj]
  {10.1086/156569}, \href
  {https://ui.adsabs.harvard.edu/abs/1978ApJ...225.1021W} {225, 1021}

\bibitem[\protect\citeauthoryear{{Woosley}}{{Woosley}}{1986}]{1986woo}
{Woosley} S.~E.,  1986, in {Audouze} J.,  {Chiosi} C.,   {Woosley} S.~E.,  eds,
  {Saas-Fee Advanced Course 16: Nucleosynthesis and Chemical Evolution}. p.~1

\bibitem[\protect\citeauthoryear{{Woosley} \& {Kasen}}{{Woosley} \&
  {Kasen}}{2010}]{2010woo}
{Woosley} S.~E.,  {Kasen} D.,  2010, preprint, \href
  {http://cdsads.u-strasbg.fr/abs/2010arXiv1010.5292W} {} (\mn@eprint {arXiv}
  {1010.5292})

\bibitem[\protect\citeauthoryear{{Woosley} \& {Weaver}}{{Woosley} \&
  {Weaver}}{1994}]{1994woob}
{Woosley} S.~E.,  {Weaver} T.~A.,  1994, \mn@doi [\apj] {10.1086/173813}, \href
  {http://cdsads.u-strasbg.fr/abs/1994ApJ...423..371W} {423, 371}

\bibitem[\protect\citeauthoryear{{Woosley}, {Arnett}  \& {Clayton}}{{Woosley}
  et~al.}{1973}]{1973woo}
{Woosley} S.~E.,  {Arnett} W.~D.,   {Clayton} D.~D.,  1973, \mn@doi [\apjs]
  {10.1086/190282}, \href {http://cdsads.u-strasbg.fr/abs/1973ApJS...26..231W}
  {26, 231}

\bibitem[\protect\citeauthoryear{{Yang} et~al.,}{{Yang} et~al.}{2018}]{2018yan}
{Yang} Y.,  et~al., 2018, \mn@doi [\apj] {10.3847/1538-4357/aa9e4c}, \href
  {https://ui.adsabs.harvard.edu/abs/2018ApJ...852...89Y} {852, 89}

\bibitem[\protect\citeauthoryear{{Zheng} et~al.,}{{Zheng}
  et~al.}{2013}]{2013zhe}
{Zheng} W.,  et~al., 2013, \mn@doi [\apj] {10.1088/2041-8205/778/1/L15}, \href
  {https://ui.adsabs.harvard.edu/abs/2013ApJ...778L..15Z} {778, L15}

\makeatother
\end{thebibliography}

% Don't change these lines
\bsp	% typesetting comment
\label{lastpage}
\end{document}